\documentclass[onecolumn,12pt,journal,final]{IEEEtran}

\makeatletter
\def\ps@headings{%
\def\@oddhead{\mbox{}\scriptsize\rightmark \hfil \thepage}%
\def\@evenhead{\scriptsize\thepage \hfil \leftmark\mbox{}}%
\def\@oddfoot{}%
\def\@evenfoot{}}
\makeatother 

\IEEEoverridecommandlockouts
\usepackage{bbm}
\usepackage{amsfonts}
\usepackage[dvips]{graphicx}
\usepackage{times}
\usepackage{cite}
\usepackage{amsmath}
\usepackage{array}
\usepackage{amssymb}

\usepackage{stfloats}
\usepackage{graphicx}
\usepackage{footnote}
\usepackage{booktabs}
\usepackage{array}
\usepackage{algorithmic}
\usepackage{algorithm}
\usepackage{subeqnarray}
\usepackage{cases}
\usepackage{threeparttable}
\usepackage{color}
\usepackage{hyperref}
\usepackage{epstopdf}
\usepackage{cleveref}
\usepackage{bm}
\usepackage{subcaption}
\usepackage{dsfont}

\newtheorem{remark}{Remark}

\newtheorem{lemma}{Lemma}

\newtheorem{claim}{Claim}

\begin{document}

\title{ADMM-based Fast Algorithm for Multi-group Multicast Beamforming in Large-Scale Wireless Systems}

\author{Erkai~Chen and Meixia~Tao\\
\thanks{This work is supported by the National Natural Science Foundation of China under grants 61571299 and 61521062. 
This paper will be presented in part at the IEEE ICC 2017 \cite{icc17submit}. 

E.~Chen and M.~Tao are with the Department of Electronic Engineering at Shanghai Jiao Tong University, Shanghai, P. R. China (emails: \{cek1006, mxtao\}@sjtu.edu.cn).
}
}

\maketitle

\begin{abstract}
Multi-group multicast beamforming in wireless systems with large antenna arrays and massive audience is investigated in this paper. Multicast beamforming design is a well-known non-convex quadratically constrained quadratic programming (QCQP) problem. A conventional method to tackle this problem is to approximate it as a semi-definite programming problem via semi-definite relaxation, whose performance, however, deteriorates considerably as the number of per-group users goes large. A recent attempt is to apply convex-concave procedure (CCP) to find a stationary solution by treating it as a difference of convex programming problem, whose complexity, however, increases dramatically as the problem size increases. In this paper, we propose a low-complexity high-performance algorithm for multi-group multicast beamforming design in large-scale wireless systems by leveraging the alternating direction method of multipliers (ADMM) together with CCP. In specific, the original non-convex QCQP problem is first approximated as a sequence of convex subproblems via CCP. Each convex subproblem is then reformulated as a novel ADMM form. Our ADMM reformulation enables that each updating step is performed by solving multiple small-size subproblems with closed-form solutions in parallel. Numerical results show that our fast algorithm maintains the same favorable performance as state-of-the-art algorithms but reduces the complexity by orders of magnitude.

\end{abstract}


\section{Introduction}
Multicasting is a promising approach to deliver a common message to multiple receivers by exploiting the broadcast nature of wireless medium. It has great potential in many applications such as live video streaming, venue casting, mobile application updates, advertisements, and public group communications \cite{Huawei_multicast, Ericsson_multicast, Qualcomm_multicast}. It can also be used in heterogeneous networks (HetNets) with wireless backhaul to push common information from a macro base station (BS) to multiple small BSs \cite{wirelessbackhaul_ICC16}. Recently, multicasting is shown to be useful for content delivery even when user demands are different in wireless cache networks \cite{fundamental_limits_caching_TIT14, Xu_ISIT16}.

Physical layer multicasting via beamforming further boosts the energy and spectrum efficiencies by exploiting channel state information at the transmitter.
Multicast beamforming is first considered in \cite{multicast06} for a single group of users. The similar problem for multiple co-channel groups is later studied in \cite{Luo_MultlGroup_multicast}. In \cite{ZhengZheng_mulitcell_multicast}, coordinated multicast beamforming among BSs in multi-cell networks is investigated. The design of multicast beamforming in cellular networks with massive multiple-input multiple-output (MIMO) is studied in \cite{Massive_MIMO_multicasting_JSAC14}.
In general, the design of multicast beamforming is a non-convex quadratically constrained quadratic programming (QCQP) problem and its global optimal solution is difficult to obtain.
A prevailing method to tackle this problem is to approximate it as a convex semi-definite programming (SDP) problem via semi-definite relaxation (SDR) \cite{multicast06,Luo_MultlGroup_multicast, ZhengZheng_mulitcell_multicast,peak-power-14}. By solving the relaxed problem, one can obtain not only a bound on the optimal performance, but also possibly a global optimal solution of the original problem if the SDP solution is rank-one. If the SDP solution is not rank-one, then it is followed by a post-processing randomization procedure to generate high-quality approximate solutions.
However, the SDR-based algorithm is not computationally efficient as it requires lifting the problem into higher dimensional space. Meanwhile, it has been demonstrated in \cite{Luo_MultlGroup_multicast} that, the performance of SDR with Gaussian randomization deteriorates considerably as the number of per-group users increases.
A more recent method to tackle the multicast beamforming problem is to apply successive convex approximation (SCA) \cite{sca1978} or convex-concave procedure (CCP) \cite{lanckriet2009convergence} to find a stationary solution \cite{multicast-sla-2014,FPP_SCA_TPL15, multigroup_multicast_FPP_SCA_SPAWC15, content_centric_TWC16}.
In particular, the CCP-based algorithms treat the problem as a difference of convex (DC) programming problem and then iteratively solve a sequence of convex subproblems constructed by replacing the concave parts of the DC functions with their first-order Taylor expansions. Both the SCA and CCP algorithms have demonstrated better performance than the SDR methods when the rank-one probability of SDP solutions is low \cite{content_centric_TWC16, multigroup_multicast_FPP_SCA_SPAWC15}.
Note that the SCA and CCP methods have also been applied recently to tackle non-convex resource allocation and beamforming design problems in various wireless networks, e.g., \cite{Tran_MASS15}, \cite{parallel-sca}.

Note that none of the aforementioned SDR or CCP algorithms for multicast beamforming design have found closed-form solutions for the approximated problems and they all rely on general-purpose solvers to find the solutions numerically. In the simplest scenario with single-group multicast beamforming, if a convex solver based on interior-point methods (e.g., CVX \cite{cvx}) is used, then the worst-case complexity is $\mathcal{O}(N^2 + K)^{3.5}$ for SDR \cite{multicast06} and is $\mathcal{O}(N + K)^{3.5}$ for each CCP iteration \cite{multicast-sla-2014}, where $N$ is the number of antennas at the transmitter and $K$ is the number of users. Clearly, the computational costs of these second-order algorithms are not scalable when the problem size increases.
Recently, ultra-dense small cell deployments \cite{CRAN_mag} and massive MIMO \cite{MassiMIMO-Mag13} have become important candidate technologies for the future generations of wireless systems (5G). They can provide high-volume and diversified data services for a large set of devices, including not only smart mobile devices operated by humans but also special-purpose machine-to-machine devices \cite{what_is_5G_JSAC14}.
The scale of these wireless systems is significantly large, dealing possibly with hundreds of antennas and users.
It is thus of particular importance to investigate low-complexity and high-performance algorithms for multicast beamforming in large-scale systems.

\subsection{Related Works}
The authors in \cite{MU-siglegroup-multicast-TSP15} propose a fast algorithm to solve the max-min fairness (MMF) problem for single-group multicast beamforming. The original problem is first approximated by replacing the MMF objective with its proportional fairness. The additive update and multiplicative update algorithms are then introduced by iteratively maximizing two different concave approximations of the new objective, both of which are updated in closed form. The algorithms are demonstrated to achieve comparable performance as CCP-based algorithms at a much lower complexity. This method, however, cannot be extended to the more general multi-group multicasting scenario.

The alternating direction method of multipliers (ADMM) \cite{SBoyd-ADMM-11} is a powerful first-order method well suited to large-scale convex optimization. It has recently been used in wireless networks for distributed or parallel computing of various resource allocation problems and beamforming design problems.
The authors in \cite{Shi_large_scale_ADMM_TSP15} present a two-stage optimization framework to efficiently solve large-scale power minimization and network utility maximization problems for dense wireless cooperative networks which can be exactly reformulated as second-order cone programming (SOCP) problems. In the two-stage approach, the original SOCP problem is first transformed into a standard cone programming form with matrix stuffing and then solved using the ADMM algorithm (i.e., the operator splitting method) proposed in \cite{SCS_2016}.
In \cite{Consensus-ADMM16}, the authors propose a consensus form of ADMM to solve a general QCQP problem. It first reformulates the QCQP problem in consensus optimization form by introducing a local copy of the optimization variables (i.e., global variable) for each single quadratic constraint and then applies ADMM to update the local and global variables alternatively. In particular, the ADMM updating step for each local copy of variables is a QCQP with only one constraint (QCQP-1) and hence can be done efficiently with possibly closed form. The authors in \cite{Consensus-ADMM16} then apply the consensus ADMM for single-group multicast beamforming problem directly. Its convergence, however, cannot be guaranteed due to the non-convexity of the multicast beamforming problem.

Both the two-stage approach for SOCPs \cite{Shi_large_scale_ADMM_TSP15} and consensus ADMM for QCQPs \cite{Consensus-ADMM16} deal with general optimization frameworks, but require a large number of auxiliary variables and hence lift the original problem into a much higher dimension space.
Customized algorithms for specific problems to enable efficient or parallel computing are developed in \cite{Multicell_beamforming_ADMM_TSP12, MingyiHong_ADMM_beamforming_TSP14} by exploiting the structures of the problems.
Specifically, in \cite{Multicell_beamforming_ADMM_TSP12}, the authors consider the robust coordinated beamforming in multi-cell networks. By using SDR approximation and S-Procedure methods, the original non-convex beamforming problem is first reformulated as a tractable convex SDP. Then an ADMM-based distributed algorithm is proposed, which is provably able to converge to the global optimum of the centralized SDP problem. The ADMM update in each BS, however, still relies on general-purpose solvers.
In \cite{MingyiHong_ADMM_beamforming_TSP14}, the joint BS activation and beamforming design for power minimization problem in HetNets is first reformulated as an SOCP using a sparsity regularizer. An efficient algorithm based on ADMM is developed to solve the SOCP, in which each updating step is in closed form and can be carried out distributively among multiple BSs.


\subsection{Contributions}
The main contribution of this work is the development of a low-complexity high-performance algorithm for multi-group multicast beamforming design by adopting the ADMM approach in conjunction with the CCP method. Both the quality-of-service (QoS) problem and MMF problems are considered, where the QoS problem is to minimize the total transmit power subject to an individual signal-to-interference-and-noise ratio (SINR) constraint for each user and a peak power constraint for each antenna, and the MMF problem is to maximize the minimum weighted SINR subject to the per-antenna peak power constraints. The proposed algorithm is first designed for the QoS problem and then extended to the MMF problem. The main technical novelty and research findings of this work are summarized as follows:
\begin{itemize}
  \item The proposed algorithm exploits the advantages of both the CCP principle and the ADMM approach. Specifically, the QoS problem in the non-convex QCQP form is first approximated with a sequence of convex subproblems by adopting CCP, which enables superior performance over the conventional SDR methods. Each convex CCP subproblem is then reformulated as a novel ADMM form that facilitates close-form solutions and parallel computing in large-scale systems.
  \item In the ADMM reformulation of each convex CCP subproblem, each updating step is decomposed into multiple small-size subproblems which are solved in parallel with closed-form solutions. Compared with the two-stage approach \cite{Shi_large_scale_ADMM_TSP15} and the consensus ADMM method \cite{Consensus-ADMM16} proposed for general frameworks, our new ADMM utilizes the specific structure of the considered multi-group multicast beamforming problem and hence requires much less auxiliary variables and reduces the complexity significantly.
  \item We further propose an efficient ADMM-based method to obtain a starting point for CCP. In the special case when the number of transmit antennas exceeds the number of users, we also find a closed-form starting point.
  \item Numerical simulations are conducted in large-scale systems with transmit antenna number $N \in [40, 250]$ and user number $K \in [50, 140]$. The results show that the proposed fast algorithm maintains the same favorable performance as the existing CCP algorithm (solved by interior-point solvers), which is within $1 \text{dB}$ close to the SDR lower bound for the QoS problem or within $0.5$dB close to the SDR upper bound for the MMF problem, but reduces the running time by orders of magnitude.
\end{itemize}

\subsection{Organization and Notations}
The rest of the paper is organized as follows. Section \ref{sec:Problem Setting} introduces the system model and problem formulations, including the QoS problem and MMF problem. Section \ref{sec:Fast Algorithm for QoS Problem} provides the details of the proposed fast algorithm for the QoS problem. The extension of the algorithm to the MMF problem is introduced in Section \ref{sec:Extension_to_MMF_Beamforming_Design}. Simulation results are provided in Section \ref{sec:simulation_results}. Finally, we conclude the paper in Section \ref{sec:conclusion}.

\emph{Notations}: In the remainder of this paper, boldface lower-case and upper-case letters denote vectors and matrices respectively. Calligraphy letters denote sets or problems, depending on the context. $\mathbb{R}$ and $\mathbb{C}$ denote the real and complex domains, respectively. $\mathbb{E}(\cdot)$ denotes the expectation of a random variable. $\mathcal{CN}(\delta,\sigma^2)$ represents a complex Gaussian distribution with mean $\delta$ and variance $\sigma^2$. The operators $(\cdot)^T$, $(\cdot)^H$, $(\cdot)^{-1}$, $(\cdot)^{\dagger}$, and $\text{Tr}(\cdot)$ correspond to the transpose, the Hermitian transpose, inverse, Moore-Penrose inverse, and trace respectively. $\lvert \cdot \rvert$ and $\lVert \cdot \rVert_2$ denote the absolute value and Euclidean norm, respectively. The real part of a complex number $x$ is denoted by $\mathfrak{R}\{x\}$. Finally, $\mathbf{I}_N$ denotes the $N \times N$ identity matrix.

\section{Problem Setting} \label{sec:Problem Setting}
We consider a multi-group multicasting system as in \cite{Luo_MultlGroup_multicast} where one transmitter, equipped with $N$ antennas, serves $M$ groups of single-antenna users. The users within each group desire a common multicast message, which is independent for different groups. Let $\mathcal{G}_m$ denote the set of users in multicast group $m$, for all $ m \in \mathcal{M} = \{1, \dots, M\}$.
Let $\mathcal{K} = \{1,\dots, K \}$ denote the set of total users in all groups. Each user participates in only one multicast group, thus $\mathcal{G}_i \cap \mathcal{G}_j = \emptyset, i \neq j, \forall i, j \in \mathcal{M}$  and $\cup_{m = 1}^M \mathcal{G}_m = \mathcal{K}$.

Let $\mathbf{w}_{m} \in \mathbb{C}^{N}$ denote the beamforming vector of group $m$, for all $ m \in \mathcal{M}$ and $\mathbf{h}_{k} \in \mathbb{C}^{N}$ denote the channel vector from the transmitter to the $k$-th user, for all $k \in \mathcal{K}$. Each channel is modeled as a complex, random vector which is flat in frequency and quasi-static in time. And it is assumed to be perfectly available at the transmitter. The corresponding received signal at user $k$ in group $m$ can be written as
\begin{equation}
y_{k} = \underbrace{\mathbf{h}_k^H \mathbf{w}_m x_m}_{\text{desired signal}} + \underbrace{\sum_{j \neq m} \mathbf{h}_k^H \mathbf{w}_j x_j}_{\text{inter-group interference}} + \underbrace{n_k}_{\text{noise}},~\forall k \in \mathcal{G}_m,
\end{equation}
where $x_m \in \mathbb{C}$ is the data symbol transmitted to multicast group $m$ with $\mathbb{E} \left[ \lvert x_m\rvert^2 \right] = 1$, and $n_k \sim \mathcal{CN}(0,\sigma_k^2)$ is the additive white Gaussian noise at user $k$.
The maximum power radiated by each antenna $n$ is denoted as $P_n$. We thus have
\begin{equation}
\sum_{m=1}^M \mathbf{w}_m^H \mathbf{R}_n \mathbf{w}_m \leq P_n,~\forall n \in \mathcal{N},
\end{equation}
where $\mathcal{N} = \{1,\dots,N\}$ is the index set of all the antennas at the transmitter and $\mathbf{R}_n \in \{0, 1\}^{N \times N}$ is the all-zero matrix except the $n$-th diagonal entry being 1, for all $n \in \mathcal{N}$. The received SINR of user $k \in \mathcal{G}_m$ is expressed as
\begin{equation}
  \text{SINR}_{k} = \frac {\lvert \mathbf{h}_k^H \mathbf{w}_m \rvert^2} {\sum_{j \neq m} \lvert \mathbf{h}_k^H \mathbf{w}_j \rvert^2 + \sigma_k^2}, ~\forall k \in \mathcal{G}_m.
\end{equation}

Similar to \cite{Luo_MultlGroup_multicast}, we consider two problem formulations for the multi-group multicast beamforming design, the QoS problem and MMF problem. The QoS problem is to minimize the total radiated power of the transmitter subject to an individual target SINR constraint for each user and a peak power constraint for each transmit antenna. It is expressed as
\begin{subequations}
\begin{align}
\mathcal{Q}:~\mathop{\text{minimize}}_{\mathbf{w} } \quad &\sum_{m=1}^M \lVert \mathbf{w}_{m} \rVert_2^2 \\
\text{subject to} \quad & \frac {\lvert \mathbf{h}_k^H \mathbf{w}_m \rvert^2} {\sum_{j \neq m} \lvert \mathbf{h}_k^H \mathbf{w}_j \rvert^2 + \sigma_k^2} \geq \gamma_k,~\forall k \in \mathcal{G}_m, \forall m \in \mathcal{M}, \label{eqn:SINR-constraint}  \\
&\sum_{m=1}^M  \mathbf{w}_{m}^H \mathbf{R}_n \mathbf{w}_{m} \leq P_n,~\forall n \in \mathcal{N}, \label{eqn:per-antenna-constraint}
\end{align} \label{eqn:QoS-problem}
\end{subequations}
where $\mathbf{w} \triangleq \{ \mathbf{w}_{m} | m \in \mathcal{M}\}$ is the set of all the beamforming vectors and $\gamma_k$ is the minimum received SINR required by user $k$.

The MMF problem is to maximize the minimum weighted SINR over all users subject to a peak power constraint for each transmit antenna. It is expressed as
\begin{subequations}
\begin{align}
\mathcal{F}:~\mathop{\text{maximize}}_{\mathbf{w},~t } \quad & t \\
\text{subject to} \quad & \frac{1}{g_k}\frac {\lvert \mathbf{h}_k^H \mathbf{w}_m \rvert^2} {\sum_{j \neq m} \lvert \mathbf{h}_k^H \mathbf{w}_j \rvert^2 + \sigma_k^2} \geq t,~\forall k \in \mathcal{G}_m, \forall m \in \mathcal{M}, \label{eqn:MMF-SINR-constraint}  \\
&\sum_{m=1}^M  \mathbf{w}_{m}^H \mathbf{R}_n \mathbf{w}_{m} \leq P_n,~\forall n \in \mathcal{N}, \label{eqn:MMF-per-antenna-constraint}
\end{align} \label{eqn:MMF-problem}
\end{subequations}
where $\{g_k\}_{k=1}^K$ is the set of weights to account for possibly different service grades among the users.

Slightly different from the original QoS problem and MMF problem formulated in \cite{Luo_MultlGroup_multicast}, our problems have included the per-antenna peak power constraints, which are more realistic in practical systems.
As mentioned in the introduction, both QoS and MMF problems can be approximately solved using the SDR method and the CCP method. In the next section, we focus on the QoS problem and propose a fast algorithm to find a  stationary solution (possibly local optimum) based on the CCP method and the ADMM approach. The extension to the MMF problem shall be briefly discussed in Section \ref{sec:Extension_to_MMF_Beamforming_Design}.

Note that while the MMF problem $\mathcal{F}$ is always feasible, the QoS problem $\mathcal{Q}$ can be infeasible if the SINR targets $\{\gamma_k \}$ and the peak power constraints $\{P_n\}$ are too stringent, or the channels of users in different multicast groups are highly correlated.
In \cite{ZhengZheng_mulitcell_multicast}, a necessary condition for the multi-cell multicast QoS beamforming problem to be feasible is given. In this paper, we derive a sufficient condition for checking the feasibility of $\mathcal{Q}$, which shall be introduced in Section \ref{Initialization of CCP algorithm}. In the following, we only discuss $\mathcal{Q}$ when it is feasible.

\section{Fast Algorithm for QoS Problem} \label{sec:Fast Algorithm for QoS Problem}
Recently, the authors in \cite{content_centric_TWC16} formulate a novel sparse multicast beamforming problem in cache-enabled cloud radio access networks which is solved using CCP-based algorithms. Since the multi-group multicast beamforming problem considered in this work is a special case of the problem in \cite{content_centric_TWC16} \footnote{The optimization problem in \cite{content_centric_TWC16} is to minimize the weighted sum of total transmission power and backhaul cost subject to a minimum SINR constraint for each multicast group. It reduces to the QoS problem considered in this work when the backhaul cost is ignored.}, the same CCP-based algorithm in \cite{content_centric_TWC16} applies here. The new contribution in this work is to develop its low-complexity implementation using ADMM. We name the proposed fast algorithm as CCP-ADMM algorithm. In this section, we first briefly outline the CCP-based algorithm, then present a novel ADMM approach to solve each CCP subproblem with closed-form solutions for the updating steps. After that, we introduce a new method to find a starting point for the CCP algorithm also based on ADMM.

\subsection{CCP Algorithm}
As in \cite{content_centric_TWC16}, the SINR constraints \eqref{eqn:SINR-constraint} in problem $\mathcal{Q}$ can be written as a DC form
\begin{equation}
	\gamma_k \left( \sum_{j \neq m} \lvert \mathbf{h}_k^H \mathbf{w}_j \rvert^2 + \sigma_k^2 \right) - \lvert \mathbf{h}_k^H \mathbf{w}_m \rvert^2 \le 0, ~\forall k \in \mathcal{G}_m, \forall m \in \mathcal{M}. \label{eqn:DC-SINR-constraint}
\end{equation}
The CCP-based algorithm is to convexify the SINR constraints in the above DC form by replacing the concave parts with their first-order Taylor expansions, then solve a sequence of convex subproblems successively. As such, CCP designed for DC programming problems is also known as a special case of SCA, which is for general non-convex problems. This iterative procedure is guaranteed to converge to a stationary point of the original problem $\mathcal{Q}$, according to \cite{lanckriet2009convergence}. Specifically, in the $t$-th iteration, we need to solve
\begin{subequations}
\begin{align}
\mathcal{Q}^{(t)}:~\mathop{\text{minimize}}_{\mathbf{w}} \quad &\sum_{m=1}^M \lVert \mathbf{w}_{m} \rVert_2^2 \\
\text{subject to} \quad &\gamma_k \left( \sum_{j \neq m} \lvert \mathbf{h}_k^H \mathbf{w}_j \rvert^2 + \sigma_k^2 \right) - 2 \mathfrak{Re} \left \{ (\mathbf{w}_m^{(t)})^H \mathbf{h}_k \mathbf{h}_k^H \mathbf{w}_m \right \} \nonumber\\
&\quad + \lvert \mathbf{h}_k^H \mathbf{w}_m^{(t)} \rvert^2  \leq 0, ~\forall k \in \mathcal{G}_m, \forall m \in \mathcal{M}, \label{eqn:convex-SINR-constraint} \\
&\sum_{m=1}^M  \mathbf{w}_{m}^H \mathbf{R}_n \mathbf{w}_{m} \leq P_n.~\forall n \in \mathcal{N}, \label{eqn:CCP-per-antenna-constraint}
\end{align} \label{eqn:CCP-iteration-problem}
\end{subequations}
where $\mathbf{w}_m^{(t)}$ is the optimal solution obtained from the previous iteration. Each $\mathcal{Q}^{(t)}$ is a convex QCQP problem and can be solved using a general-purpose solver through interior-point methods. In what follows, we exploit the specific structure of problem $\mathcal{Q}^{(t)}$ and find its optimal solution using an ADMM-based fast algorithm.

Note that a parallel SCA algorithm is proposed in \cite{parallel-sca} to solve a resource allocation problem in a multi-cell cellular system with coordinated multi-point transmission (CoMP).
In each iteration of the parallel SCA, all BSs are enabled to update their variables through parallelly solving multiple convexified and decoupled subproblems, one for each BS. Different from \cite{parallel-sca}, where the optimization variables from different BSs are separated in the constraints, the optimization variables $\{\mathbf{w}_m\}$ of our problem are coupled in the constraints (i.e., \eqref{eqn:SINR-constraint} and \eqref{eqn:per-antenna-constraint}). We could use the similar idea to approximate the original problem $\mathcal{Q}$ by a sequence of parallel subproblems with each $\mathbf{w}_m$ decoupled in both the objective and the constraints. However, the solutions of these parallel subproblems in each SCA iteration may not be feasible to the original problem $\mathcal{Q}$ as the originally coupled constraints may no longer be satisfied.

\subsection{ADMM algorithm for each CCP subproblem}
We first introduce two sets of auxiliary variables (their significance will be discussed in the end of this subsection):
\begin{align}
\Gamma_{k, m} &= \mathbf{h}_k^H \mathbf{w}_m,~\forall k \in \mathcal{K}, m \in \mathcal{M}, \\
\mathbf{v}_m &= \mathbf{w}_m,~\forall m \in \mathcal{M},
\end{align}
where $\Gamma_{k, m}$ represents the interference level at user $k$ caused by the beamforming vector of group $m$, and $\mathbf{v}_m \in \mathbb{C}^N $ is a copy of the original beamforming vector $\mathbf{w}_m$. To ease the notation, we define $\bm{\Gamma} \triangleq \{ \Gamma_{k, m} | k \in \mathcal{K}, m \in \mathcal{M}\}$, and $\mathbf{v} \triangleq \{ \mathbf{v}_{m} | m \in \mathcal{M}\}$.

Then problem $\mathcal{Q}^{(t)}$ can be equivalently expressed as
\begin{subequations}
\begin{align}
\mathop{\text{minimize}}_{\{\bm{\Gamma}, \mathbf{v}, \mathbf{w}\}} \quad &\sum_{m=1}^M \lVert \mathbf{w}_{m} \rVert_2^2 \\
\text{subject to} \quad & \Gamma_{k, m} - \mathbf{h}_k^H \mathbf{w}_m = 0,~\forall k \in \mathcal{K}, m \in \mathcal{M}, \label{eqn:Gamma-consistent-constraint}\\
& \mathbf{v}_m - \mathbf{w}_m = 0,~\forall m \in \mathcal{M}, \label{eqn:V-consistent-constraint}\\
& \gamma_k \left ( \sum_{j \neq m} \lvert \Gamma_{k, j} \rvert^2 + \sigma_k^2 \right ) - 2 \mathfrak{Re} \left \{ (\mathbf{w}_m^{(t)})^H \mathbf{h}_k \Gamma_{k, m} \right \} \nonumber \\
&\quad + \lvert \mathbf{h}_k^H \mathbf{w}_m^{(t)} \rvert^2 \leq  0, ~\forall k \in \mathcal{G}_m, \forall m \in \mathcal{M}, \label{eqn:CCP-admm-SINR-constraint}  \\
& \sum_{m=1}^M \mathbf{v}_{m}^H \mathbf{R}_n \mathbf{v}_{m} \leq P_n,~\forall n \in \mathcal{N}.  \label{eqn:CCP-admm-perantenna-constraint}
\end{align} \label{eqn:CCP-admm-subproblem}
\end{subequations}
We define the feasible region of constraint \eqref{eqn:CCP-admm-SINR-constraint} as $ \mathcal{C}$, and its indicator function as
\begin{equation}
\mathbb{I}_{\mathcal{C}} (\bm{\Gamma}) =
\begin{cases}
	0, &\text{ if } \bm{\Gamma} \in \mathcal{C}, \\
	+\infty, &\text{ otherwise}.	
\end{cases}
\end{equation}
Similarly, we define the feasible region of constraint \eqref{eqn:CCP-admm-perantenna-constraint} as $ \mathcal{D}$, and its indicator function as
\begin{equation}
\mathbb{I}_{\mathcal{D}} (\mathbf{v}) =
\begin{cases}
	0, &\text{ if } \mathbf{v} \in \mathcal{D}, \\
	+\infty, &\text{ otherwise}.	
\end{cases}
\end{equation}
Then, we obtain the equivalent ADMM reformulation of problem $\mathcal{Q}^{(t)}$
\begin{align} \label{eqn:admm-problem}
\mathop{\text{minimize}}_{\{\bm{\Gamma}, \mathbf{v}, \mathbf{w}\}} \quad &\sum_{m=1}^M \lVert \mathbf{w}_{m} \rVert_2^2  + \mathbb{I}_{\mathcal{C}} (\bm{\Gamma}) + \mathbb{I}_{\mathcal{D}} (\mathbf{v})\\
\text{subject to} \quad & \eqref{eqn:Gamma-consistent-constraint},\eqref{eqn:V-consistent-constraint}. \nonumber
\end{align}
The augmented Lagrangian (using the scaled dual variable) of problem \eqref{eqn:admm-problem} is given by
\begin{align}
\mathcal{L}_{\rho} (\bm{\Gamma}, \mathbf{v}, \mathbf{w}, \bm{\lambda}, \mathbf{z}) =&  \sum_{m=1}^M \lVert \mathbf{w}_{m} \rVert_2^2  + \mathbb{I}_{\mathcal{C}} (\bm{\Gamma}) + \mathbb{I}_{\mathcal{D}} (\mathbf{v}) + \frac{\rho}{2} \sum_{k=1}^K \sum_{m=1}^M \lvert \Gamma_{k, m} - \mathbf{h}_k^H \mathbf{w}_m + \lambda_{k,m} \rvert^2  \nonumber \\ & \quad + \frac{\rho}{2} \sum_{m=1}^M \lVert \mathbf{v}_m - \mathbf{w}_m + \mathbf{z}_{m} \rVert_2^2, \label{eqn:Lagrangian-admm-problem}
\end{align}
where $\rho > 0$ is the penalty parameter, $\bm{\lambda} \triangleq \{ \lambda_{k, m} | k \in \mathcal{K}, m \in \mathcal{M}\}$ and $\mathbf{z} \triangleq \{ \mathbf{z}_{m} \in \mathbb{C}^N | m \in \mathcal{M}\}$ are the scaled dual variables for constraints \eqref{eqn:Gamma-consistent-constraint} and \eqref{eqn:V-consistent-constraint}, respectively.

From problem \eqref{eqn:admm-problem}, we observe that the variables in the constraints can be split into two blocks, $ \{\mathbf{\Gamma}, \mathbf{v} \}$ and $\mathbf{w}$, and that the objective function is also separable across this splitting.
Thus, by adopting ADMM, we can minimize $ \mathcal{L}_{\rho} (\bm{\Gamma}, \mathbf{v}, \mathbf{w}, \bm{\lambda}, \mathbf{z})$ by updating the two blocks of variables, $\{\mathbf{\Gamma}, \mathbf{v} \}$ and $\mathbf{w}$, alternatively. The ADMM procedure is given in Alg. \ref{alg:QoS-subproblem-ADMM}.

\begin{algorithm}[h]
\caption{ADMM for solving problem $\mathcal{Q}^{(t)}$} \label{alg:QoS-subproblem-ADMM}
\begin{algorithmic}[0]
\STATE \textbf{Initialization:} Initialize $\mathbf{w}_{m}^{0} \leftarrow \mathbf{w}_{m}^{(t)}, \mathbf{z}_{m}^{0} \leftarrow \mathbf{0}, \lambda_{k,m}^{0} \leftarrow 0, \forall m \in \mathcal{M}, k \in \mathcal{K}$, and $j \leftarrow 0$. Set the penalty parameter $\rho$.
\STATE \textbf{Repeat}
\begin{enumerate}
	\item Update the first block of variables $\{\mathbf{\Gamma}^{j+1}, \mathbf{v}^{j+1} \}$
   	\begin{align} \label{eqn:admm-Gamma-V-update}
	\{\mathbf{\Gamma}^{j+1}, \mathbf{v}^{j+1}\}:= \arg\min_{\mathbf{\Gamma}, \mathbf{v}} \mathcal{L}_{\rho} (\bm{\Gamma}, \mathbf{v}, \mathbf{w}^{j}, \bm{\lambda}^{j}, \mathbf{z}^{j}).
	\end{align}
	\item Update the second block of variables $\mathbf{w}^{j+1}$
	\begin{align} \label{eqn:admm-W-update}
	\mathbf{w}^{j+1} := \arg\min_{\mathbf{w}} \mathcal{L}_{\rho} (\bm{\Gamma}^{j+1}, \mathbf{v}^{j+1}, \mathbf{w}, \bm{\lambda}^{j}, \mathbf{z}^{j}).
	\end{align}
	\item Update the dual variables $\{\bm{\lambda}^{j+1}, \mathbf{z}^{j+1} \}$
	\begin{align}
	&\lambda^{j+1}_{k,m} := \lambda^j_{k,m} + (\Gamma_{k, m}^{j+1} - \mathbf{h}_k^H \mathbf{w}_m^{j+1}),~\forall k \in \mathcal{K}, m \in \mathcal{M}, \label{eqn:admm-lambda-update}\\
	&\mathbf{z}^{j+1}_{m} := \mathbf{z}^j_{m} + (\mathbf{v}_{m}^{j+1} - \mathbf{w}_m^{j+1}),~\forall m \in \mathcal{M}. \label{eqn:admm-delta-update}
	\end{align}
   \item Set $j \leftarrow j + 1$.
\end{enumerate}
\STATE \textbf{Until} convergence criterion is met.
\end{algorithmic}
\end{algorithm}

In the following, we elaborate the details of updating the primal variables $\{\mathbf{\Gamma}, \mathbf{v}\}$ and $\mathbf{w}$ (the superscript ADMM iteration counter is ignored for simplicity). Note that the update for the first block of variables in \eqref{eqn:admm-Gamma-V-update} can be decomposed into two independent problems, one for each of the two sets $ \mathbf{\Gamma} $ and $\mathbf{v}$, which are given in \eqref{eqn:admm-Gamma-update} and \eqref{eqn:admm-V-update}, respectively.

\begin{align} \label{eqn:admm-Gamma-update}
\mathbf{\Gamma}^{j+1} := \arg\min_{\mathbf{\Gamma}} \left \{ \mathbb{I}_{\mathcal{C}} (\mathbf{\Gamma}) + \frac{\rho}{2} \sum_{k=1}^K \sum_{m=1}^M \lvert \Gamma_{k, m} - \mathbf{h}_k^H \mathbf{w}_m^{j} + \lambda_{k,m}^{j}\rvert^2 \right \}
\end{align}
\begin{align} \label{eqn:admm-V-update}
\mathbf{v}^{j+1} := \arg\min_{\mathbf{v}} \left \{\mathbb{I}_{\mathcal{D}} (\mathbf{v}) + \frac{\rho}{2} \sum_{m=1}^M \lVert \mathbf{v}_{m} - \mathbf{w}_m^{j} + \mathbf{z}_{m}^{j}\rVert_2^2 \right \}
\end{align}

\subsubsection{$\mathbf{\Gamma}$ Update} \label{sec:CCP-ADMM_Gamma_update}
The update of $\mathbf{\Gamma}$ in problem \eqref{eqn:admm-Gamma-update} is equivalent to solving the problem:
\begin{align} \label{eqn:admm-Gamma-update-problem}
\mathop{\text{minimize}}_{\bm{\Gamma}} \quad &\sum_{k=1}^K \sum_{m=1}^M \lvert \Gamma_{k, m} - \mathbf{h}_k^H \mathbf{w}_m^{j} + \lambda_{k,m}^{j}\rvert^2 \\
\text{subject to} \quad &\eqref{eqn:CCP-admm-SINR-constraint}. \nonumber
\end{align}
It is observed that problem \eqref{eqn:admm-Gamma-update-problem} can be decomposed into $K$ subproblems, one for each $k \in \mathcal{K}$:
\begin{subequations}
\begin{align}
\mathop{\text{minimize}}_{\{\Gamma_{k,m}\}_{m=1}^M} \quad & \sum_{m=1}^M \lvert \Gamma_{k, m} - \mathbf{h}_k^H \mathbf{w}_m^{j} + \lambda_{k,m}^{j}\rvert^2 \\
\text{subject to}\quad & \gamma_k \left ( \sum_{m \neq m_k} \lvert \Gamma_{k, m} \rvert^2 + \sigma_k^2 \right ) - 2 \mathfrak{Re} \left \{ (\mathbf{w}_{m_k}^{(t)})^H \mathbf{h}_k \Gamma_{k, m_k} \right \} + \lvert \mathbf{h}_k^H \mathbf{w}_{m_k}^{(t)} \rvert^2 \leq  0, \label{eqn:Gamma-constraint}
\end{align} \label{eqn:admm-Gamma-update-subproblem}
\end{subequations}
where $m_k$ is the index of the multicast group that user $k$ belongs to. The superscript $j$ and $(t)$ are the ADMM and CCP iteration counters, respectively. We note that each subproblem \eqref{eqn:admm-Gamma-update-subproblem} is a convex QCQP-1 problem. In various cases, QCQP-1 problems can be solved efficiently \cite{Consensus-ADMM16}. In our case, the optimal solution of subproblem \eqref{eqn:admm-Gamma-update-subproblem} is obtained in closed form. The details are given Appendix \ref{solution-for-admm-Gamma-update-subproblem}.

\subsubsection{$\mathbf{v}$ Update} \label{sec:CCP-ADMM_v_update}
The update of $\mathbf{v}$ in problem \eqref{eqn:admm-V-update} is equivalent to solving the problem:
\begin{align} \label{eqn:admm-V-update-problem}
\mathop{\text{minimize}}_{ \mathbf{v} } \quad & \sum_{m=1}^M \lVert \mathbf{v}_{m} - \mathbf{w}_m^{j} + \mathbf{z}_{m}^{j}\rVert_2^2 \\
\text{subject to}\quad & \eqref{eqn:CCP-admm-perantenna-constraint}.  \nonumber
\end{align}
By inspection, this problem can be decomposed into $N$ subproblems, one for each antenna $n \in \mathcal{N}$:
\begin{align} \label{eqn:admm-V-update-subproblem}
\mathop{\text{minimize}}_{ \{v_{n,m}\}_{m=1}^M } \quad & \sum_{m=1}^M \lvert v_{n, m} - w_{n, m}^{j} + z_{n, m}^{j}\rvert^2 \\
\text{subject to}\quad & \sum_{m=1}^M \lvert v_{n, m} \rvert^2 \leq P_n, \nonumber
\end{align}
where $v_{n,m}$ is the $n$-th element of vector $\mathbf{v}_m$, $w_{n,m}$ and $z_{n,m}$ are defined in a similar manner.
Subproblem \eqref{eqn:admm-V-update-subproblem} is also a QCQP-1 problem. We solve this problem optimally in closed form. In specific, we rewrite subproblem \eqref{eqn:admm-V-update-subproblem} as
\begin{align} \label{eqn:admm-V-update-projection-subproblem}
\mathop{\text{minimize}}_{ \tilde{\mathbf{v}}_{n} } \quad & \lVert \tilde{\mathbf{v}}_{n} - \tilde{\mathbf{u}}_{n}^{j}\rVert_2 \\
\text{subject to}\quad & \lVert \tilde{\mathbf{v}}_{n} \rVert_2 \leq \sqrt{P_n}, \nonumber
\end{align}
where $\tilde{\mathbf{v}}_{n} = [v_{n, 1}, \dots, v_{n, M}]^T \in \mathbb{C}^{M}$ and $\tilde{\mathbf{u}}_{n}^{j} = [(w_{n, 1}^{j} - z_{n, 1}^{j}), \dots, (w_{n, M}^{j} - z_{n, M}^{j})]^T \in \mathbb{C}^{M}$. It is clear that problem \eqref{eqn:admm-V-update-projection-subproblem} can be viewed as the Euclidean projection of the point $\tilde{\mathbf{u}}_{n}^{j}$ onto an Euclidean ball, centered at the original point with radius of $\sqrt{P_n}$. The optimal solution is thus given by
\begin{align} \label{eqn:admm-V-update-solution}
\tilde{\mathbf{v}}_{n} = \min \left \{\frac{\sqrt{P_n}}{\lVert \tilde{\mathbf{u}}_{n}^{j} \rVert_2}, 1 \right \} \tilde{\mathbf{u}}_{n}^{j}.
\end{align}
Here we give an intuitive geometric explanation for the solution \eqref{eqn:admm-V-update-solution}. If the point $\tilde{\mathbf{u}}_{n}^{j}$ is already inside the ball (i.e., $\lVert \tilde{\mathbf{u}}_{n}^{j} \rVert_2 \leq \sqrt{P_n}$), then the point $\tilde{\mathbf{u}}_{n}^{j}$ itself is what we want. Otherwise, we simply scale it to have Euclidean norm equal to $\sqrt{P_n}$.

\subsubsection{$\mathbf{w}$ Update}
The problem \eqref{eqn:admm-V-update} for updating the second block of variables $\mathbf{w}$ can be decomposed into $M$ independent unconstrained quadratic programing problems, one for each group $m \in \mathcal{M}$:
\begin{align} \label{eqn:admm-W-update-problem}
\mathbf{w}_m^{j+1} := \arg\min_{\mathbf{w}_{m}} \Bigg \{ & \lVert \mathbf{w}_{m} \rVert_2^2 + \frac{\rho}{2} \sum_{k=1}^K  \lvert \Gamma_{k, m}^{j+1} - \mathbf{h}_k^H \mathbf{w}_m + \lambda_{k,m}^{j}\rvert^2 + \frac{\rho}{2} \lVert \mathbf{v}_{m}^{j+1} - \mathbf{w}_m + \mathbf{z}_{m}^{j}\rVert_2^2 \Bigg \}.
\end{align}
The solution is given in closed form as
\begin{align} \label{eqn:admm-W-update-solution}
\mathbf{w}_{m}^{j+1} = \left((2 + \rho) \mathbf{I}_N + \rho \sum_{k} \mathbf{h}_k \mathbf{h}_k^H \right)^{-1} \left ( \rho\sum_{k} \mathbf{h}_k ( \Gamma^{j+1}_{k,m} + \lambda_{k,m}^j ) + \rho (\mathbf{v}_m^{j+1} + \mathbf{z}_{m}^{j} )\right).
\end{align}
Note that the most computational intensive operation in Alg. \ref{alg:QoS-subproblem-ADMM} is the matrix inversion in $\mathbf{w}$ update \eqref{eqn:admm-W-update-solution}, whose complexity is $\mathcal{O}(N^3)$. However, this operation only needs to be computed once for each channel realization and the solution can be readily used in the subsequent iterations.

Up to now, the closed-form expressions for all the updating steps in Alg. \ref{alg:QoS-subproblem-ADMM} have been derived. The ADMM iteration can converge to a global optimum of problem \eqref{eqn:CCP-admm-subproblem} \cite[Proposition 4.2]{bertsekas1997parallel}. Note that Alg. \ref{alg:QoS-subproblem-ADMM} is guaranteed to converge for any initial point. We propose to initialize $\mathbf{w}_m^{0}$ using the solution obtained in the previous CCP iteration (i.e., $\mathbf{w}_m^{(t)}$) in Alg. \ref{alg:QoS-subproblem-ADMM}. It is a warm start and can often speed up the convergence.


\begin{remark}
The main novelty of our proposed ADMM algorithm lies in the design of the auxiliary variables $\{\Gamma_{k, m}\}$ and $\{\mathbf{v}_m\}$, with which,  problem $\mathcal{Q}^{(t)}$ can be reformulated into such a form that each updating step in ADMM is decomposable and thus can be updated through parallelly solving multiple subproblems with much smaller sizes. Specifically, the variables $\{\Gamma_{k, m}\}$ enable that the SINR constraints in \eqref{eqn:convex-SINR-constraint} with coupled variables is transformed into the new constraint in \eqref{eqn:CCP-admm-SINR-constraint} with decoupled variables. As a result, problem \eqref{eqn:admm-Gamma-update-problem} can be decomposed into $K$ QCQP-1 subproblems and then solved efficiently in closed form. The introduction of variables $\{\mathbf{v}_m\}$ makes it possible that the $\mathbf{w}$ update in \eqref{eqn:admm-W-update} can be decomposed into $M$ small-scale unconstrained problems and carried out in closed form. Meanwhile, the update of $\{\mathbf{v}_m\}$ itself can also be efficiently performed through an Euclidean projection onto an Euclidean ball given in \eqref{eqn:admm-V-update-solution}.
\end{remark}

\begin{remark} \label{remark_conADMM}
The consensus ADMM proposed in \cite{Consensus-ADMM16} can be applied to optimally solve problem $\mathcal{Q}^{(t)}$ too. The details are given in Appendix \ref{Consensus-ADMM Implementation}. However, since a local copy of the global optimization variables $\mathbf{w}$ is introduced for each single quadratic constraint in \eqref{eqn:convex-SINR-constraint} and \eqref{eqn:CCP-per-antenna-constraint}, yielding a total of $KMN + MN^2$ auxiliary variables, the problem size of each local variable update is still as large as the original problem $\mathcal{Q}^{(t)}$. In comparison, the number of introduced auxiliary variables in our proposed ADMM in Alg. \ref{alg:QoS-subproblem-ADMM} is only $KM+NM$, which is much smaller. This is especially useful when the number of antennas $N$ is large. Numerical simulations also demonstrate the complexity advantage of our proposed ADMM.
One may also apply the consensus ADMM in \cite{Consensus-ADMM16} to directly solve the original QoS problem $\mathcal{Q}$ in \eqref{eqn:QoS-problem} without using CCP. The iteration, however, is not guaranteed to converge due to the non-convexity of problem $\mathcal{Q}$. Numerical results in Section \ref{sec:simulation_results} show that such method performs poorly for the considered multi-group multicast beamforming problem.
\end{remark}

\subsection{Initialization of CCP Algorithm} \label{Initialization of CCP algorithm}
The CCP-based algorithms generally need a feasible starting point, which is difficult to obtain in general.
In \cite{content_centric_TWC16}, the authors propose to initialize the CCP with a feasible point found by SDR with Gaussian randomization. This method works well in simulation, though the extra complexity of SDR method is high.
In \cite{Consensus-ADMM16}, the authors find a feasible point for initialization using the same consensus ADMM for the general QCQP problems. In this work, we find a starting (not necessary feasible due to the per-antenna peak power constraints) point for initialization of the CCP-ADMM algorithm also using ADMM.

We first formulate a feasibility problem that takes	into account the non-convex SINR constraints \eqref{eqn:SINR-constraint} only and ignores the per-antenna peak power constraints \eqref{eqn:per-antenna-constraint} as:
\begin{subequations}
\begin{align}
\mathop{\text{find}}  \quad &\{\bm{\Gamma}, \mathbf{w}\} \\
\text{such that} \quad & \Gamma_{k, m} - \mathbf{h}_k^H \mathbf{w}_m = 0,~\forall k \in \mathcal{K}, m \in \mathcal{M}, \\
& \gamma_k \left ( \sum_{j \neq m} \lvert \Gamma_{k, j} \rvert^2 + \sigma_k^2 \right ) - \lvert \Gamma_{k, m}\rvert^2 \leq  0, ~\forall k \in \mathcal{G}_m, \forall m \in \mathcal{M}. \label{eqn:FPP-admm-SINR-constraint}
\end{align} \label{eqn:FPP-admm-subproblem}
\end{subequations}
By applying ADMM, the iterative steps are given by
\begin{align}
\mathbf{\Gamma}^{j+1} := &\arg\min_{\mathbf{\Gamma}} \left \{ \sum_{k=1}^K \sum_{m=1}^M \lvert \Gamma_{k, m} - \mathbf{h}_k^H \mathbf{w}_m^{j} + \lambda_{k,m}^{j}\rvert^2 \right \} \label{eqn:FPP-admm-Gamma-update} \\
&\quad \quad \text{s.t.}~\eqref{eqn:FPP-admm-SINR-constraint},  \nonumber\\
\mathbf{w}^{j+1} := &\arg\min_{\mathbf{w}} \left \{ \sum_{k=1}^K \sum_{m=1}^M \lvert \Gamma_{k, m}^{j+1} - \mathbf{h}_k^H \mathbf{w}_m + \lambda_{k,m}^{j}\rvert^2 + \sum_{m=1}^M \lVert \mathbf{v}_{m}^{j+1} - \mathbf{w}_m + \mathbf{z}_{m}^{j}\rVert_2^2 \right \}, \label{eqn:FPP-admm-W-update} \\
\lambda^{j+1}_{k,m} := &\lambda^j_{k,m} + (\Gamma_{k, m}^{j+1} - \mathbf{h}_k^H \mathbf{w}_m^{j+1}),~\forall k \in \mathcal{K}, m \in \mathcal{M}, \label{eqn:FPP-admm-lambda-update}
\end{align}
which are independent of the penalty parameter $\rho$.

The $\mathbf{\Gamma}$ update in \eqref{eqn:FPP-admm-Gamma-update} is similar to \eqref{eqn:admm-Gamma-update}, which can also be decomposed into $K$ subproblems, one for each $k \in \mathcal{K}$:
\begin{subequations}
\begin{align}
\mathop{\text{minimize}}_{\{\Gamma_{k,m}\}_{m=1}^M} \quad & \sum_{m=1}^M \lvert \Gamma_{k, m} - \mathbf{h}_k^H \mathbf{w}_m^{j} + \lambda_{k,m}^{j}\rvert^2 \\
\text{subject to}\quad & \gamma_k \left ( \sum_{m \neq m_k} \lvert \Gamma_{k, m} \rvert^2 + \sigma_k^2 \right ) - \lvert  \Gamma_{k, m_k} \rvert^2 \leq  0. \label{eqn:FPP-admm-Gamma-update-constraint}
\end{align} \label{eqn:FPP-admm-Gamma-update-subproblem}
\end{subequations}
Each subproblem \eqref{eqn:FPP-admm-Gamma-update-subproblem} is still a QCQP-1 problem, but non-convex, in contrast to \eqref{eqn:admm-Gamma-update-subproblem} which is convex. Nevertheless, the strong duality still holds, and problem \eqref{eqn:FPP-admm-Gamma-update-subproblem} can be solved optimally irrespective of the non-convexity \cite[Appendix B]{Boyd_convex_optimization}. Its closed-form solution is given in Appendix \ref{solution-for-FPP-admm-Gamma-update-subproblem}.

The update of $\mathbf{w}$ in problem \eqref{eqn:FPP-admm-W-update} is given by
\begin{align} \label{eqn:FPP-admm-W-update-solution}
\mathbf{w}_{m}^{j+1} = \left(\mathbf{I}_N + \sum_{k} \mathbf{h}_k \mathbf{h}_k^H \right)^{-1} \left (\sum_{k} \mathbf{h}_k ( \Gamma^{j+1}_{k,m} + \lambda_{k,m}^j ) +  (\mathbf{v}_m^{j+1} + \mathbf{z}_{m}^{j} )\right),~\forall m \in \mathcal{M}.
\end{align}

With random initialization, the ADMM iteration terminates until a point, denoted as $\mathbf{w}^{(0)}$ that satisfies the SINR constraints is found. This type of iteration usually converges very fast, if such a point $\mathbf{w}^{(0)}$ exists. Note that, with different initializations, the ADMM iteration may find different $\mathbf{w}^{(0)}$. If it fails to find the point even after a large number of trials, then we have, to some extent, numerical evidence that the problem may be infeasible. Once $\mathbf{w}^{(0)}$ is obtained, it can be used as a starting point for the CCP at iteration $t = 0$.

\begin{remark}
The obtained starting point $\mathbf{w}^{(0)}$ is not necessarily a {\em feasible} point of the original QoS problem in \eqref{eqn:QoS-problem} since it may not satisfy the per-antenna peak power constraints \eqref{eqn:per-antenna-constraint}. However, the per-antenna peak power constraints will automatically be satisfied by the subsequent $\mathbf{w}^{(t)}$ for all iterations $t \geq 1$ because of the constraint \eqref{eqn:CCP-per-antenna-constraint}. One might find a truly {\em feasible} starting point using the same ADMM by including the per-antenna peak power constraints \eqref{eqn:per-antenna-constraint} into the above feasibility problem \eqref{eqn:FPP-admm-subproblem}. Our numerical results, however, suggest that such ADMM method can hardly find a feasible point with stringent per-antenna power constraints, even though the feasible set is non-empty.
\end{remark}
\begin{remark}
It might be possible that the problem $\mathcal{Q}^{(t)}$ for $t = 0$ is infeasible at the obtained starting point $\mathbf{w}^{(0)}$. If this happens, we try multiple $\mathbf{w}^{(0)}$ until this problem is feasible. Note that multiple $\mathbf{w}^{(0)}$ can be obtained by using ADMM iteration \eqref{eqn:FPP-admm-Gamma-update}-\eqref{eqn:FPP-admm-lambda-update} with different initializations.
\end{remark}

In a special case, we can find an alternative starting point with closed-form expression, as given in the following lemma.
\begin{lemma} \label{lemma_feasible_point}
If the aggregate channel matrix $\mathbf{H} = [\mathbf{h}_{1},\dots,\mathbf{h}_{K}] \in \mathbb{C}^{N \times K}$ has full column rank, then there always exists a starting point that satisfies the SINR constraints \eqref{eqn:SINR-constraint} and it is given by
\begin{align} \label{eqn:solution-W}
\mathbf{W} = [\mathbf{w}_1, \mathbf{w}_2, \dots, \mathbf{w}_M] = \mathbf{H}(\mathbf{H}^H\mathbf{H})^{-1} \mathbf{A},
\end{align}
where $\mathbf{A}$ is a $K \times M$ complex-valued matrix with the $(k, m)$-th element defined as
\begin{align}  \label{eqn:solution-A}
A_{k, m} =
\begin{cases}
	\sqrt{\gamma_k \sigma_k^2}  e^{j \theta_{k}}, &\text{ if } k \in \mathcal{G}_m, \\
	0, &\text{ otherwise}.
\end{cases}
\end{align}
where $\theta_{k} \in [0, 2\pi]$ is arbitrary. Further, if the starting point in \eqref{eqn:solution-W} satisfies the per-antenna peak power constraints \eqref{eqn:per-antenna-constraint}, it is also a feasible solution of problem $\mathcal{Q}$.
\end{lemma}

\begin{IEEEproof}
See Appendix \ref{proof_of_lemma_feasible_point}.
\end{IEEEproof}

This lemma guarantees that a starting point that satisfies the SINR constraints exists with probability one and has a closed-form expression if $N \geq K$ and the entries of each $\mathbf{h}_{k}$ are independent and identically distributed. The lemma also provides a sufficient condition for checking the feasibility of problem $\mathcal{Q}$.

\section{Extension to MMF Beamforming Design}
\label{sec:Extension_to_MMF_Beamforming_Design}
In this section, we extend the CCP-ADMM algorithm proposed for the QoS problem to the MMF problem.

\subsection{Duality of MMF Problem $\mathcal{F}$}
In \cite{Luo_MultlGroup_multicast}, without the per-antenna peak power constraints, the authors prove that the QoS problem and MMF problem are a dual pair. Including the per-antenna peak power constraints, a similar duality is established in \cite{peak-power-14}. We restate the conclusions in \cite{peak-power-14} as follows. Consider a per-antenna power minimization problem \cite{peak-power-14}
\begin{subequations}
\begin{align}
\mathcal{P}:~\mathop{\text{minimize}}_{\mathbf{w},~r } \quad & r \\
\text{subject to} \quad & \frac {\lvert \mathbf{h}_k^H \mathbf{w}_m \rvert^2} {\sum_{j \neq m} \lvert \mathbf{h}_k^H \mathbf{w}_j \rvert^2 + \sigma_k^2} \geq g_k,~\forall k \in \mathcal{G}_m, \forall m \in \mathcal{M}, \label{eqn:MMF-QoS-SINR-constraint}  \\
&\frac{1}{P_n} \sum_{m=1}^M \mathbf{w}_{m}^H \mathbf{R}_n \mathbf{w}_{m} \leq r,~\forall n \in \mathcal{N}. \label{eqn:MMF-QoS-per-antenna-constraint}
\end{align} \label{eqn:MMF-QoS-minimization-problem}
\end{subequations}
Following the same notation in the literature \cite{Luo_MultlGroup_multicast,peak-power-14}, let $\mathbf{g} = [g_1, \dots, g_K]^T \in \mathbb{R}^{K}$ be the user target SINR weight vector and $\mathbf{p} = [P_1, \dots, P_N]^T \in \mathbb{R}^{N}$ be the per-antenna peak power vector. Problems $\mathcal{F}$ and $\mathcal{P}$ receive $\mathbf{g}$ and $\mathbf{p}$ as inputs and the optimal objective values are denoted as $t^{\star} = \mathcal{F}(\mathbf{g}, \mathbf{p})$ and $r^{\star} = \mathcal{P}(\mathbf{g}, \mathbf{p})$, respectively. Then we have the following claims:

\begin{claim}[Claim 2 in \cite{peak-power-14}] \label{clm:F-P-relation}
Problems $\mathcal{F}$ and $\mathcal{P}$ are related as
\begin{align}
& 1 = \mathcal{P}(\mathcal{F}(\mathbf{g},  \mathbf{p})\cdot \mathbf{g}, \mathbf{p}), \label{MMF-QoS-relation-1}\\
& t = \mathcal{F}(\mathbf{g}, \mathcal{P}(t \cdot \mathbf{g}, \mathbf{p}) \cdot \mathbf{p}). \label{MMF-QoS-relation-2}
\end{align}
\end{claim}

\begin{claim} \label{clm:nondecraesing-P}
The optimum objective value of problem $\mathcal{P}(t \cdot \mathbf{g}, \mathbf{p})$ is monotonically nondecreasing in $t$, for a given $\mathbf{g}$ and $\mathbf{p}$.
\end{claim}

{\em Proof:} The feasible region of problem $\mathcal{P}(t \cdot \mathbf{g}, \mathbf{p})$ becomes smaller as $t$ increases, thus completing the proof. \hfill $\Box$

By the above claims, a solution to the MMF problem $\mathcal{F}(\mathbf{g}, \mathbf{p})$ can be found by iteratively solving $\mathcal{P}(t \cdot \mathbf{g}, \mathbf{p})$ through adjusting the value of $t$. Claim \ref{clm:F-P-relation} guarantees the optimality of the solution for $1 = \mathcal{P}(t^{\star} \cdot \mathbf{g}, \mathbf{p})$ and Claim \ref{clm:nondecraesing-P} enables the use of a simple one-dimensional bisection search for the sought $t$. Given the non-negativity of $t$, a lower bound of $t$ can be set as $L = 0$. An upper bound of $t$ can be obtained by transmitting all the available power $P_{\text{all}} = \sum_{n = 1}^N P_n$ towards the user with the best channel condition, namely, $U = \max_{k \in \mathcal{K}} \{(P_{\text{all}} \|\mathbf{h}_k \|_2^2) / (g_k \sigma_k^2)\}$.
	
\subsection{Fast Algorithm for Solving  Problem $\mathcal{P}$}
We first introduce the following auxiliary variables:
\begin{align}
\Gamma_{k, m} &= \mathbf{h}_k^H \mathbf{w}_m,~\forall k \in \mathcal{K}, m \in \mathcal{M}, \\
\mathbf{v}_m &= \mathbf{w}_m,~\forall m \in \mathcal{M}, \\
\alpha_n &= r,~\forall n \in \mathcal{N},
\end{align}
where $\alpha_{n}$ represents the local copy of power weight $r$ for the $n$-th antenna. To facilitate the notation, we define $\bm{\alpha} \triangleq \{ \alpha_{n} | n \in \mathcal{N}\}$.

By adopting CCP, in each iteration, we solve the following problem $\mathcal{P}^{(t)}$ with ADMM reformulation
\begin{subequations}
\begin{align}
\mathcal{P}^{(t)}:~\mathop{\text{minimize}}_{\{\bm{\Gamma}, \mathbf{v}, \mathbf{w}, \bm{\alpha}, r\}} \quad &r \\
\text{subject to} \quad & \Gamma_{k, m} - \mathbf{h}_k^H \mathbf{w}_m = 0,~\forall k \in \mathcal{K}, m \in \mathcal{M}, \\
& \mathbf{v}_m - \mathbf{w}_m = 0,~\forall m \in \mathcal{M}, \\
& \mathbf{\alpha}_n - r = 0,~\forall n \in \mathcal{N}, \label{equ:MMF-delta-r-constraint}\\
& g_k \left ( \sum_{j \neq m} \lvert \Gamma_{k, j} \rvert^2 + \sigma_k^2 \right ) - 2 \mathfrak{Re} \left \{ (\mathbf{w}_m^{(t)})^H \mathbf{h}_k \Gamma_{k, m} \right \} \nonumber \\
&\quad + \lvert \mathbf{h}_k^H \mathbf{w}_m^{(t)} \rvert^2 \leq  0, ~\forall k \in \mathcal{G}_m, \forall m \in \mathcal{M},  \label{eqn:MMF-CCP-admm-SINR-constraint} \\
& \frac{1}{P_n} \sum_{m=1}^M \mathbf{v}_{m}^H \mathbf{R}_n \mathbf{v}_{m} \leq \alpha_n,~\forall n \in \mathcal{N}.  \label{eqn:MMF-CCP-admm-perantenna-constraint}
\end{align} \label{eqn:MMF-CCP-admm-subproblem}
\end{subequations}
To solve problem $\mathcal{P}^{(t)}$ using ADMM, the two blocks of variables $\{\mathbf{\Gamma}, \mathbf{v}, \bm{\alpha}\}$ and $\{\mathbf{w}, r\}$ are alternatively updated. In particular, the ADMM updates (scaled form) for problem \eqref{eqn:MMF-CCP-admm-subproblem} are given by Alg. \ref{alg:MMF-subproblem-ADMM}, where $\bm{\mu} \triangleq \{\mu_n | n \in \mathcal{N} \}$ is the dual variable for constraints \eqref{equ:MMF-delta-r-constraint}.

\begin{algorithm}[ht]
\caption{ADMM for solving problem $\mathcal{P}^{(t)}$} \label{alg:MMF-subproblem-ADMM}
\begin{algorithmic}[0]
\STATE \textbf{Initialization:} Initialize $\mathbf{w}_{m}^{0} \leftarrow \mathbf{w}_{m}^{(t)}, \mathbf{z}_{m}^{0} \leftarrow \mathbf{0}, \lambda_{k,m}^{0} \leftarrow 0, \mu_{n}^{0} \leftarrow 0, \forall m \in \mathcal{M}, k \in \mathcal{K}, n \in \mathcal{N}, r^{0} \leftarrow \max_{n \in \mathcal{N}} \left \{\frac{1}{P_n} \sum_{m=1}^M (\mathbf{w}_{m}^{(t)})^H \mathbf{R}_n \mathbf{w}_{m}^{(t)}\right \}$, and $j \leftarrow 0$. Set the penalty parameter $\rho$.
\STATE \textbf{Repeat}
\begin{enumerate}
	\item Update the first block of variables $\{\mathbf{\Gamma}^{j+1}, \mathbf{v}^{j+1}, \bm{\alpha}^{j+1} \}$
	\begin{align}
	\mathbf{\Gamma}^{j+1} := \arg\min_{\mathbf{\Gamma}} &\left \{ \sum_{k=1}^K \sum_{m=1}^M \lvert \Gamma_{k, m} - \mathbf{h}_k^H \mathbf{w}_m^{j} + \lambda_{k,m}^{j}\rvert^2 \right \} \label{eqn:MMF-admm-Gamma-update}\\
	\text{s.t.}~  &\eqref{eqn:MMF-CCP-admm-SINR-constraint}, \nonumber
	\end{align}
	\begin{align}
	\{\mathbf{v}^{j+1}, \bm{\alpha}^{j+1}\} := \arg\min_{\mathbf{v}, \bm{\alpha}} &\sum_{m=1}^M \lVert \mathbf{v}_{m} - \mathbf{w}_m^{j} + \mathbf{z}_{m}^{j}\rVert_2^2 + \frac{\rho}{2} \sum_{n=1}^N \lvert \alpha_n - r^{j} + \mu_{n}^{j}\rvert^2 \label{eqn:MMF-admm-V-delta-update} \\
	\text{s.t.}~  &\eqref{eqn:MMF-CCP-admm-perantenna-constraint}. \nonumber
	\end{align}
	\item Update the second block of variables $\{\mathbf{w}^{j+1}, r^{j+1}\}$
	\begin{align}
	\mathbf{w}^{j+1} := \arg\min_{\mathbf{w}} \left \{ \sum_{k=1}^K \sum_{m=1}^M \lvert \Gamma_{k, m}^{j+1} - \mathbf{h}_k^H \mathbf{w}_m + \lambda_{k,m}^{j}\rvert^2 + \sum_{m=1}^M \lVert \mathbf{v}_{m}^{j+1} - \mathbf{w}_m + \mathbf{z}_{m}^{j}\rVert_2^2 \right \}, \label{eqn:MMF-admm-W-update}
	\end{align}
	\begin{align}
	r^{j+1} := \arg\min_{r} \left \{ r +  \frac{\rho}{2} \sum_{n=1}^N  \lvert \alpha_{n}^{j+1} - r + \mu_{n}^{j}\rvert^2 \right \}. \label{eqn:MMF-admm-r-update}
	\end{align}
	\item Update the dual variables $\{\bm{\lambda}^{j+1}, \mathbf{z}^{j+1}, \bm{\mu}^{j+1} \}$
	\begin{align}
	&\lambda^{j+1}_{k,m} := \lambda^j_{k,m} + (\Gamma_{k, m}^{j+1} - \mathbf{h}_k^H \mathbf{w}_m^{j+1}),~\forall k \in \mathcal{K}, m \in \mathcal{M}, \label{eqn:MMF-admm-lambda-update}\\
	&\mathbf{z}^{j+1}_{m} := \mathbf{z}^j_{m} + (\mathbf{v}_{m}^{j+1} - \mathbf{w}_m^{j+1}),~\forall m \in \mathcal{M}, \label{eqn:MMF-admm-mu-update} \\
	&\mu^{j+1}_{n} := \mu^j_{n} + (\alpha_n^{j+1} - r^{j+1}),~\forall n \in \mathcal{N}. \label{eqn:MMF-admm-alpha-update}
	\end{align}
   \item Set $j \leftarrow j + 1$.
\end{enumerate}
\STATE \textbf{Until} convergence criterion is met.
\end{algorithmic}
\end{algorithm}

Note that the updates of $\mathbf{\Gamma} $ and $\{ \mathbf{v}, \bm{\alpha} \}$ can be separated, both of which can be further decomposed into multiple QCQP-1 subproblems with much smaller size, and performed in closed form. The procedure for $\{ \mathbf{v}, \bm{\alpha} \}$ update is similar to $\mathbf{\Gamma}$ update in Section \ref{sec:CCP-ADMM_Gamma_update}, thus is omitted here.
The updates of $\mathbf{w}$ in \eqref{eqn:MMF-admm-W-update} and $r$ in \eqref{eqn:MMF-admm-r-update} can be obtained in closed form through solving two unconstrained quadratic programs, which are given by \eqref{eqn:FPP-admm-W-update-solution}
and \eqref{eqn:MMF-admm-r-update-solution}, respectively.
\begin{align} \label{eqn:MMF-admm-r-update-solution}
r = \frac{1}{N} \sum_{n=1}^N (\alpha_{n}^{j+1} + \mu_{n}^{j}) - \frac{1}{N \rho}.
\end{align}

A starting point $\mathbf{w}^{(0)}$ satisfying the SINR constraints can be obtained as suggested in Section \ref{Initialization of CCP algorithm}. Together with $r^{(0)} = \max_{n \in \mathcal{N}} \frac{1}{P_n} \sum_{m=1}^M (\mathbf{w}^{(0)}_{m})^H \mathbf{R}_n \mathbf{w}^{(0)}_{m}$, $\{\mathbf{w}^{(0)}, r^{(0)}\}$ is a feasible point of the original problem $\mathcal{P}$.

\subsection{On the Extension to NUM Problems}
Our proposed optimization framework can be extended to general network utility maximization (NUM) problems in the multicast transmission scenario but not necessarily in an optimal way.
Similar to the unicast scenario as in \cite[Section II-C]{Shi_large_scale_ADMM_TSP15}, the NUM problem in multicast transmission is also a monotonic optimization problem for a general strictly increasing utility (e.g., weighted sum-rate) and thus can be solved using the polyblock outer approximation algorithm or the branch-reduce-and-bound algorithm through solving a series of feasibility subproblems with given SINR targets \cite{bjornson2013optimal}. Different from unicast transmission, the feasibility subproblems with given SINR targets in the multicast transmission are non-convex in general. However, we can still apply the proposed ADMM to determine the feasibility of the subproblems numerically and approximately.

\section{Simulation Results} \label{sec:simulation_results}
In this section, the performance and complexity of our fast algorithm CCP-ADMM are demonstrated via numerical simulations. The following baselines are selected,
\begin{itemize}
	\item \textbf{SDR-GauRan:} the SDR method with Gaussian randomization proposed in \cite{Luo_MultlGroup_multicast}, but modified slightly to cope with the per-antenna peak power constraints. The relaxed SDP problem is solved using the CVX package via interior-point solver SDPT3. If the solution is not rank-one, $200$ Gaussian randomization samples are adopted.
	\item \textbf{FPP-SCA:} the feasible point pursuit SCA algorithm proposed in \cite{FPP_SCA_TPL15,multigroup_multicast_FPP_SCA_SPAWC15}, which is equivalent to the CCP algorithm in \cite{content_centric_TWC16} except the initialization. The convex subproblems are solved using CVX via SDPT3 solver.
  	\item \textbf{ConADMM:} the consensus ADMM in \cite{Consensus-ADMM16} that is directly applied to solve the original non-convex QCQP problem without using CCP as mentioned in Remark \ref{remark_conADMM}. The penalty parameter is set as $\rho = 30$ to prevent $\{ \mathbf{w}^j \}$ from diverging towards infeasibility. If the ADMM iteration cannot converge to a feasible point within $3000$ iterations, a scaling operation similar to \cite{Luo_MultlGroup_multicast} is then followed to refine the ADMM solution. If the scaling problem is still infeasible, we claim that ConADMM fails.
  	\item \textbf{CCP-ConADMM:} the CCP method with each subproblem solved using the consensus ADMM in \cite{Consensus-ADMM16} as mentioned in Remark \ref{remark_conADMM} with details given in Alg. \ref{alg:QoS-subproblem-consensus-ADMM}.
  	\item \textbf{CCP-CVX-SCS:} the CCP method with each subproblem first transformed into the standard cone programming form via CVX then solved using the general ADMM proposed in \cite{Shi_large_scale_ADMM_TSP15,SCS_2016} via SCS solver. The maximum iteration and the convergence tolerance are set to be $3000$ and $10^{-6}$, respectively. Other settings are the same as in \cite{SCS_2016}.
\end{itemize}

For the CCP (or SCA)-based algorithms (namely, FPP-SCA, CCP-ConADMM, CCP-CVX-SCS, CCP-ADMM), the iteration stops when the relative decrease of the objective $\frac{\lvert p^{(t+1)} - p^{(t)} \rvert} {p^{(t)}}$ is less than $10^{-3}$ or a maximum of $30$ iterations is reached, where $p^{(t)}$ denotes the objective value of the $t$-th iteration.
For CCP-ADMM and CCP-ConADMM, the penalty parameters in Alg. \ref{alg:QoS-subproblem-ADMM} and Alg. \ref{alg:QoS-subproblem-consensus-ADMM} are set as $\rho = \frac{2}{\sqrt{N}}$ and $\rho = \frac{10}{\sqrt{N}}$, respectively, which are empirically found to work very well (converge fast).
The convergence criterion of the ADMM is set as suggested in \cite[Section 3.3]{SBoyd-ADMM-11} with the absolute tolerance $\epsilon^{\text{abs}} = 10^{-6}$ and relative tolerance $\epsilon^{\text{rel}} = 10^{-6}$. If the ADMM fails to converge within $3000$ iterations, we claim the problem is infeasible.

The downlink channels of all users are assumed to be independent and follow the standard complex Gaussian distribution $\mathcal{CN}(0, 1)$. The noise variance is set to $\sigma_k^2 = 1, \forall k$. The SINR target for each user is $\gamma_k = 10 \text{dB}, \forall k$.
All experiments are carried out on a Windows x64 machine with 3.3 GHz CPU and 24 GB of RAM. The plots are obtained after averaging over $100$ channel realizations.
The performance of our fast algorithm is mainly validated for the QoS problem as similar conclusions can be extended to the MMF problem.

\subsection{Convergence Analysis} \label{Convergence-Analysis}
\begin{figure}[tbp]
\begin{centering}
\includegraphics[scale=.30]{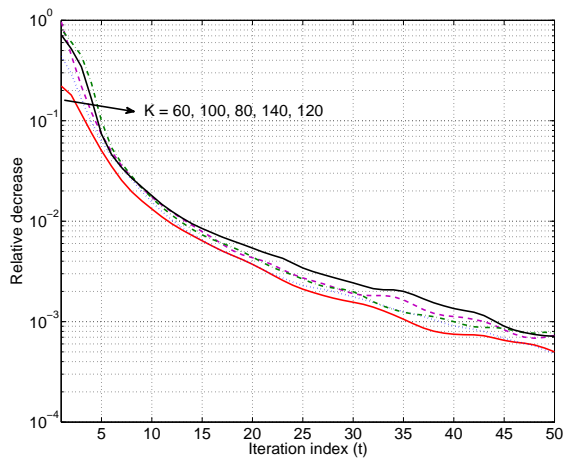}
 \caption{\small{Convergence behavior of the CCP algorithm in the outer loop $(N = 100)$. }}\label{fig:fig_convergence_CCP}
\end{centering}
\end{figure}

\begin{figure}[tbp]
\begin{centering}
\includegraphics[scale=.30]{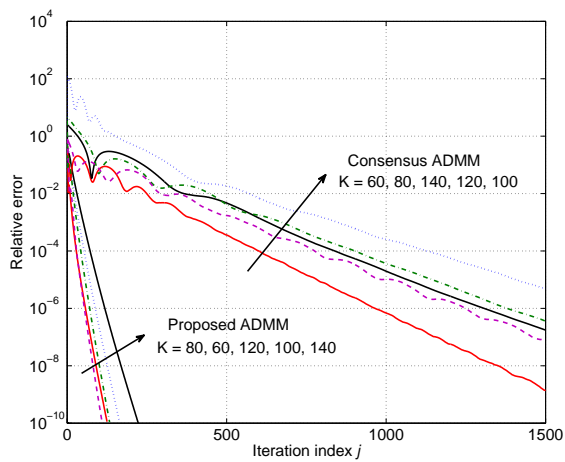}
 \caption{\small{Convergence behavior of the ADMM algorithm in the inner loop $(N = 100)$. }}\label{fig:fig_convergence_ADMM}
\end{centering}
\end{figure}

We first demonstrate the convergence behavior of the proposed CCP-ADMM algorithm. The algorithm involves an outer-loop iteration for CCP and an inner-loop iteration for ADMM, whose convergences are illustrated in Fig.~\ref{fig:fig_convergence_CCP} and Fig.~\ref{fig:fig_convergence_ADMM}, respectively. Here, in Fig.~\ref{fig:fig_convergence_ADMM}, the relative error in each iteration is defined as $\frac{\lvert a^{j} - a^* \rvert}{a^*}$, where $a^{j}$ is the objective value of the $j$-th iteration and $a^*$ is the optimal objective value of the given problem $\mathcal{Q}^{(t)}$  \footnote{ The optimal objective value $a^*$ can be obtained using interior-point methods.}. The number of transmit antennas is fixed as $N = 100$, and the number of users $K$ varies from $60$ to $140$. All the users are equally divided into $M = 4$ multicast groups. The maximum transmit power for each antenna is set to $P_n = 40 \text{dBm}, \forall n \in \mathcal{N}$. From Fig.~\ref{fig:fig_convergence_CCP}, it is observed that the CCP algorithm converges to modest accuracy, e.g., relative decrease of $10^{-2}$ within $15$ iterations and then higher accuracy as the iterations progress. From Fig.~\ref{fig:fig_convergence_ADMM}, it is seen that the proposed ADMM algorithm converges to a high accuracy solution, e.g., relative error of $10^{-4}$, within $100$ iterations for various $K$, while the existing consensus ADMM needs about 700-1200 iterations to achieve the same level of accuracy.

\subsection{Performance and Complexity Comparison}
\subsubsection{Comparison with varying number of users}

\begin{figure}[tbp]
\begin{centering}
\includegraphics[scale=.30]{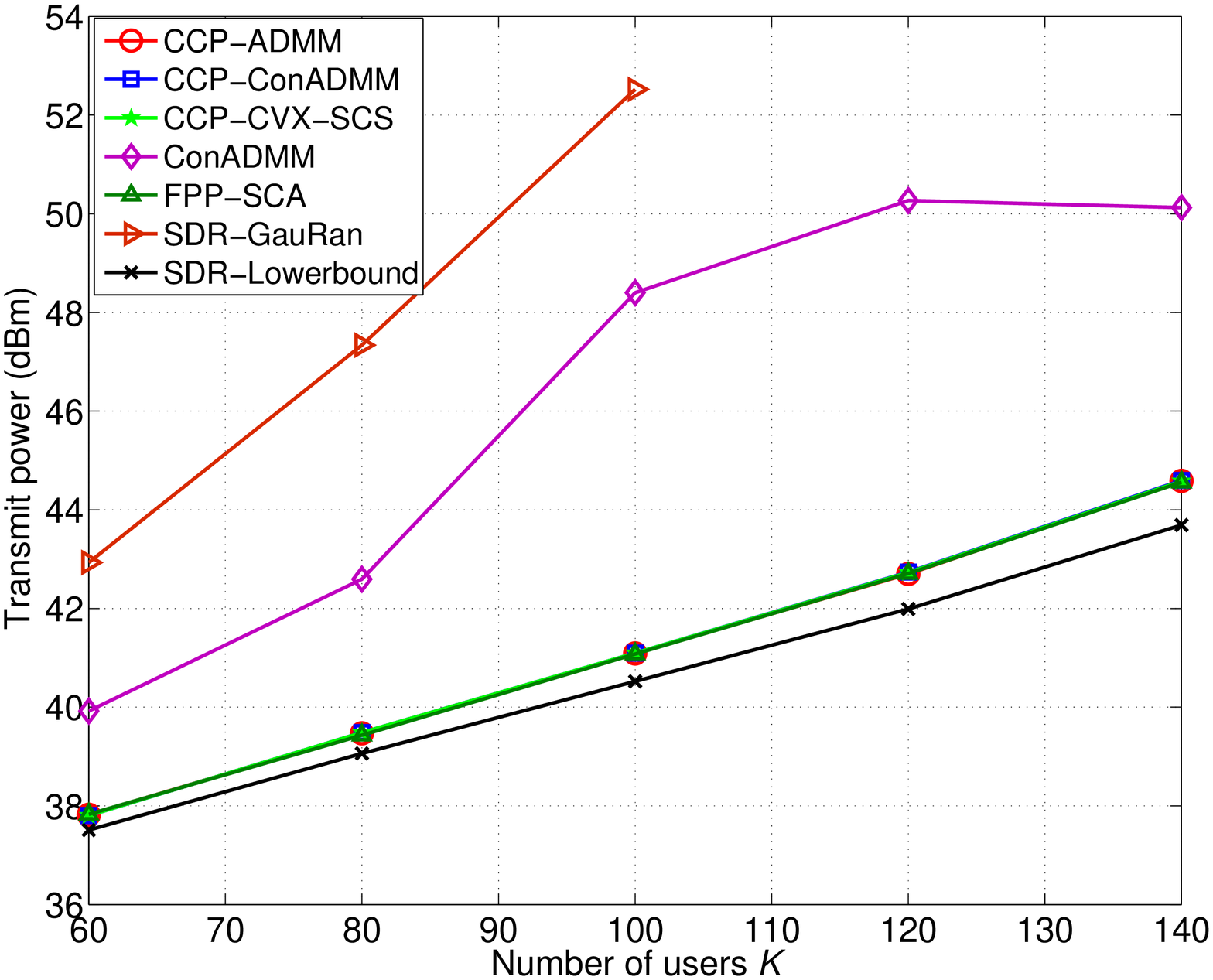}
 \caption{\small{Transmit power versus the number of users $K$ ($N = 100$). }}\label{fig:fig_power_user}
\end{centering}
\end{figure}

\begin{figure}[tbp]
\begin{centering}
\includegraphics[scale=.30]{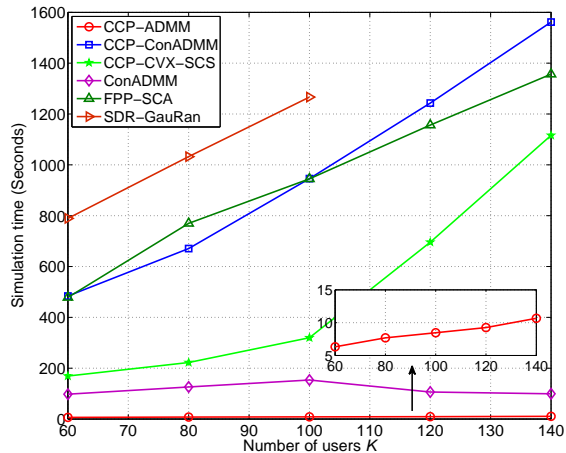}
 \caption{\small{Simulation time versus the number of users $K$ ($N = 100$). }}\label{fig:fig_time_user}
\end{centering}
\end{figure}

With the same scenario as in Section~\ref{Convergence-Analysis}, Fig.~\ref{fig:fig_power_user} and Fig.~\ref{fig:fig_time_user} compare the performance in terms of the actual transmit power and the computational complexity in terms of the actual simulation running time, respectively.
We first observe that the conventional SDR-GauRan method can hardly find a feasible solution where $K > 100$, and its gap to the SDR lower bound is as large as $12 \text{dB}$ when $K = 100$. The reason is that the rank-one probability of SDP solutions is very low in the considered scenario. We also see that ConADMM performs much worse than our proposed CCP-ADMM in terms of power efficiency and simulation running time. This is mainly due to that ConADMM fails to converge to a feasible solution and the refinement must be performed most of the time.

From Fig.~\ref{fig:fig_power_user}, it is also seen that the three CCP-based algorithms, namely, CCP-ADMM, CCP-ConADMM, and CCP-CVX-SCS, achieve the same favorable performance as FPP-SCA, as expected, and their solutions are within $1 \text{dB}$ close to the SDR lower bound. Their computational complexities are however different. It is seen from Fig.~\ref{fig:fig_time_user} that the running time of CCP-ADMM is about $70\sim 170$ times faster than FPP-SCA and CCP-ConADMM. It is also $30\sim 100$ times faster than CCP-CVX-SCS \footnote{ The running time of CCP-CVX-SCS includes the modeling time for transforming the original problem instance into the standard form using CVX and the solving time for calling SCS solver.}.
This is mainly because the consensus ADMM in \cite{Consensus-ADMM16} and the general ADMM in \cite{Shi_large_scale_ADMM_TSP15,SCS_2016} need to introduce a large number of auxiliary variables to transform the original problem instance into certain standard forms, the dimension of which are much larger than the original problem. In addition, the consensus ADMM numerically converges much slower than our proposed ADMM, as shown in Fig.~\ref{fig:fig_convergence_ADMM}. The modeling time for transforming the original problem instance into the standard form in CCP-CVX-SCS is also non-negligible.

We would like to remark that since the consensus ADMM in \cite{Consensus-ADMM16} and the general ADMM in \cite{Shi_large_scale_ADMM_TSP15,SCS_2016} deal with general frameworks, they are not really comparable to our proposed tailor-made algorithm. We mainly use them to verify the high-performance and low-complexity of our customized algorithm in terms of the solution quality and the running time via numerical simulation.

\subsubsection{Comparison with varying number of transmit antennas}
\begin{figure}[tbp]
\begin{centering}
\includegraphics[scale=.30]{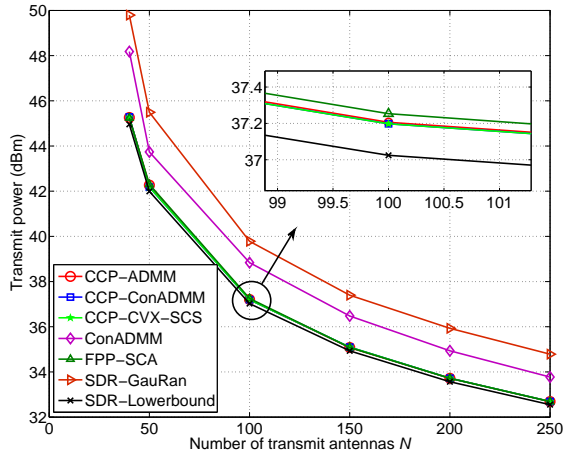}
 \caption{\small{Transmit power versus the number of transmit antennas $N$ ($K = 50$). }}\label{fig:fig_power_antenna}
\end{centering}
\vspace{-0.1cm}
\end{figure}

\begin{figure}[tbp]
\begin{centering}
\includegraphics[scale=.30]{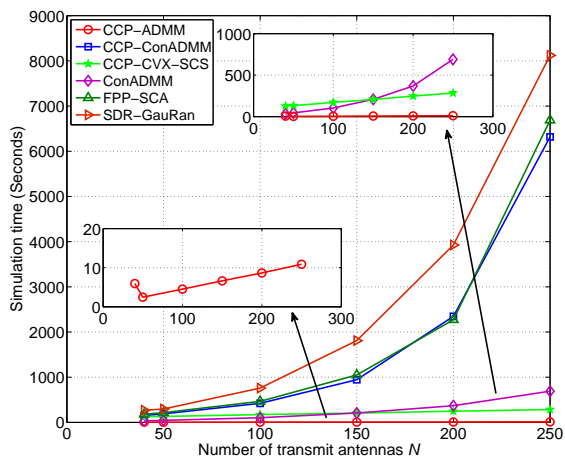}
 \caption{\small{Simulation time versus the number of transmit antennas $N$ ($K = 50$). }}\label{fig:fig_time_antenna}
\end{centering}
\vspace{-0.1cm}
\end{figure}

The performance and complexity comparison at $K=50$ users with varying number of transmit antennas are illustrated in Fig.~\ref{fig:fig_power_antenna} and Fig.~\ref{fig:fig_time_antenna}, respectively.
Here, the users are equally divided into $M = 5$ multicast groups. The maximum transmit power for each antenna is set as $P_n = P_{\text{all}}/N, \forall n \in \mathcal{N}$,  where $P_{\text{all}} = 57 \text{dBm}$ ($500$ W). This ensures that the total transmit power keeps constant when the number of antennas increases.

Similar observations as in the previous two figures can be observed. In particular, the CCP-ADMM achieves the same performance as FPP-SCA, which is within $0.3 \text{dB}$ close to the SDR lower bound at several orders of magnitude reduction in complexity. It is also observed from Fig.~\ref{fig:fig_time_antenna} that as the number of transmit antennas increases, the timing curves of CCP-ADMM and CCP-CVX-SCS grow almost linearly, and the curves of the other four baselines grow almost exponentially. Compared with CCP-CVX-SCS, our proposed CCP-ADMM can speedup the running time about $30\sim 50$ times. Note that the discontinuity of the running time of CCP-ADMM at $N=40$ is due to that the closed-form starting point \eqref{eqn:solution-W} in Lemma~\ref{lemma_feasible_point} only holds for $N \geq K (=50)$ and hence the starting point at $N = 40$ has to be found using the ADMM iterations \eqref{eqn:FPP-admm-Gamma-update}-\eqref{eqn:FPP-admm-lambda-update}, which is more time-consuming.

\subsubsection{Comparison for the MMF problem}

\begin{figure}[tbp]
\begin{centering}
\includegraphics[scale=.30]{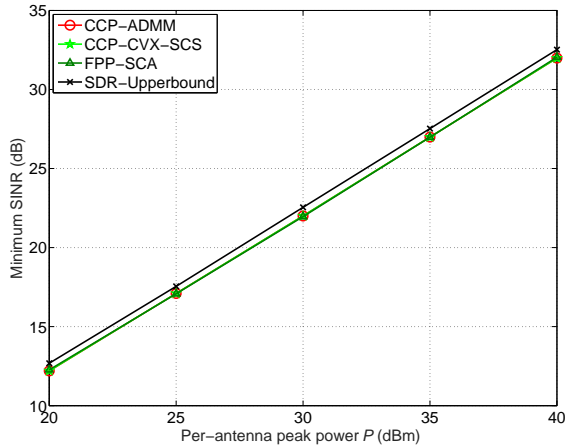}
 \caption{\small{Minimum SINR versus the per-antenna peak power $P$ ($P_n = P,~\forall n$).}}\label{fig:fig_rate_power}
\end{centering}
\vspace{-0.2cm}
\end{figure}

\begin{table}[tbp]
\caption{Average simulation time of the MMF problem.}  \label{tab:simulation-time-MMF}
\centering
\begin{tabular}{|c|c|c|c|}
\hline
Algorithm & CCP-ADMM & CCP-CVX-SCS & FPP-SCA \\
\hline
Time (sec) & 71.53 & 898.23 & 905.39 \\
\hline
\end{tabular}
\vspace{-0.2cm}
\end{table}

We finally demonstrate the performance of our fast algorithm for the MMF problem. The obtained minimum SINR (with equal weights) at different per-antenna peak power levels is illustrated in Fig.~\ref{fig:fig_rate_power}. The corresponding average simulation running time is shown in Table~\ref{tab:simulation-time-MMF}. Here, $N=100$ transmit antennas and $K = 50$ users evenly divided into $M = 5$ groups.
It is seen that our proposed CCP-ADMM achieves the same performance as CCP-CVX-SCS and FPP-SCA, which is within $0.5$dB close to the SDR upper bound, but the running time is more than $10$ times faster.

The above numerical results verify that the proposed CCP-ADMM has high performance and extremely low complexity compared with the state-of-the-art algorithms in the literature.

\section{Conclusion}
\label{sec:conclusion}
This paper develops a fast algorithm for the multi-group multicast beamforming problem in large-scale wireless systems. The algorithm adopts the CCP method to maintain near-optimal performance and utilizes the ADMM to realize low-complex and parallel implementation.
In addition, we propose an efficient ADMM-based method to obtain a starting point for the algorithm. Numerical results verify that the proposed fast algorithm reduces the complexity by multiple orders of magnitude compared with the state-of-the-art algorithms and is very suitable for large-scale wireless systems.

\appendices
\section{Finding a closed-form solution of Problem \eqref{eqn:admm-Gamma-update-subproblem}} \label{solution-for-admm-Gamma-update-subproblem}
The Lagrangian of problem \eqref{eqn:admm-Gamma-update-subproblem} is given by
\begin{align}
\mathcal{L}_{\pi} =& \sum_{m=1}^M \lvert \Gamma_{k, m} - \mathbf{h}_k^H \mathbf{w}_m^{j} + \lambda_{k,m}^{j}\rvert_2^2 \nonumber\\
&\quad + \pi  \left [ \gamma_k \left ( \sum_{m \neq m_k} \lvert \Gamma_{k, m} \rvert^2 + \sigma_k^2 \right ) - 2 \mathfrak{Re} \left \{ (\mathbf{w}_{m_k}^{(t)})^H \mathbf{h}_k \Gamma_{k, m_k} \right \} + \lvert \mathbf{h}_k^H \mathbf{w}_{m_k}^{(t)} \rvert^2  \right ], \label{eqn: Lagrangian-admm-Gamma-update-subproblem}
\end{align}
where $\pi \geq 0$ is the dual variable associated with the inequality constraint \eqref{eqn:Gamma-constraint}. By setting the derivative of $\mathcal{L}_{\pi}$ with respect to $\{\Gamma_{k,m} \}_{m=1}^M$ to zero, i.e., $ \partial \mathcal{L}_{\pi} / \partial \Gamma_{k, m} = 0, \forall m \in \mathcal{M}$, the optimal primal variables $\{\Gamma_{k,m} \}_{m=1}^M$ can be expressed as
\begin{align}  \label{equ:admm-Gamma-solution}
\Gamma_{k, m} =
\begin{cases}
	( \mathbf{h}_k^H \mathbf{w}_m^{j} - \lambda_{k,m}^{j} ) / (\pi \gamma_k + 1) ,&\text{ if } m \neq m_k, m \in \mathcal{M}, \\
	\pi \mathbf{h}_k^H \mathbf{w}_m^{(t)} + \mathbf{h}_k^H \mathbf{w}_m^{j} - \lambda_{k,m}^{j}, &\text{ if } m = m_k.
\end{cases}
\end{align}
Next, we find the optimal dual variable $\pi^*$. Substituting the solution \eqref{equ:admm-Gamma-solution} back into the inequality constraint \eqref{eqn:Gamma-constraint}, we have
\begin{align}
g(\pi) = &  \frac{ \gamma_k \sum_{m \neq m_k} \lvert \mathbf{h}_k^H \mathbf{w}_m^{j} - \lambda_{k,m}^{j} \rvert^2} {(\pi \gamma_k + 1)^2} - 2 \pi \lvert \mathbf{h}_k^H \mathbf{w}_{m_k}^{(t)} \rvert^2 \nonumber \\
&\quad + \gamma_k\sigma_k^2 - 2 \mathfrak{Re} \left \{ (\mathbf{w}_{m_k}^{(t)})^H \mathbf{h}_k (\mathbf{h}_k^H \mathbf{w}_{m_k}^{j} - \lambda_{k,m_k}^{j})\right \} + \lvert \mathbf{h}_k^H \mathbf{w}_{m_k}^{(t)} \rvert^2  \leq 0. \nonumber
\end{align}

Taking a closer look at the function $g(\pi)$, we obtain its derivative
\begin{align}
\nabla g(\pi) = - \frac{ 2 \gamma_k^2 \sum_{m \neq m_k} \lvert \mathbf{h}_k^H \mathbf{w}_m^{j} - \lambda_{k,m}^{j} \rvert^2} {(\pi \gamma_k + 1)^3} - 2  \lvert \mathbf{h}_k^H \mathbf{w}_{m_k}^{(t)} \rvert^2 \nonumber < 0. \nonumber
\end{align}
Obviously, there is $\nabla g(\pi) < 0$ (Note that when $\nabla g(\pi) = 0 $, problem \eqref{eqn:admm-Gamma-update-subproblem} is infeasible.). Thus, $g(\pi)$ is monotonically decreasing in the region $\pi \geq 0$. According to the complementary slackness condition, we have $\pi^* g(\pi^*) = 0$. Therefore, if $g(0) < 0$, we have $\pi^* = 0$. Otherwise, $g(\pi)$ has a zero-crossing in the region $\pi \geq 0$, and we have $g(\pi^*) = 0$. Thus, $\pi^*$ can be found efficiently using a root finding method, such as bisection search or Newton’s method \cite{Consensus-ADMM16}. In our context, closed-form solution of $\pi^*$ can be derived. For ease of notation, we let
\begin{align}
&a \triangleq \gamma_k  \sum_{m \neq m_k} \lvert \mathbf{h}_k^H \mathbf{w}_m^{j} - \lambda_{k,m}^{j} \rvert^2, \nonumber \\
&b \triangleq -2 \lvert \mathbf{h}_k^H \mathbf{w}_{m_k}^{(t)} \rvert^2, \nonumber \\
&c \triangleq \gamma_k\sigma_k^2 - 2 \mathfrak{Re} \left \{ (\mathbf{w}_{m_k}^{(t)})^H \mathbf{h}_k (\mathbf{h}_k^H \mathbf{w}_{m_k}^{j} - \lambda_{k,m_k}^{j})\right \} + \lvert \mathbf{h}_k^H \mathbf{w}_{m_k}^{(t)} \rvert^2. \nonumber
\end{align}
Then solving $g(\pi) = 0$ is equivalent to solving the cubic equation
\begin{align}
h(\pi) = (b\gamma_k^2) \pi^3 + (2b\gamma_k + c \gamma_k^2)\pi^2 + (b + 2 c \gamma_k)\pi + (c + a) = 0,
\end{align}
of which the closed-form solution can be expressed using the cubic formula. The real non-negative root is guaranteed to be the unique (thus correct) solution for $\pi^*$. Then, the optimal $\{\Gamma_{k,m}\}_{m=1}^M$ can be recovered by substituting $\pi^*$ into equation \eqref{equ:admm-Gamma-solution}.

\section{Consensus-ADMM Algorithm for Problem $\mathcal{Q}^{(t)}$} \label{Consensus-ADMM Implementation}
By adopting the consensus form of ADMM in \cite{Consensus-ADMM16}, we introduce a auxiliary variable copy of $\mathbf{w}$ for every single quadratic constraint, and problem $\mathcal{Q}^{(t)}$ can be equivalently expressed as
\begin{subequations}
\begin{align}
\mathop{\text{minimize}}_{\{\mathbf{x}_{k}, \mathbf{y}_{n}, \mathbf{w}\}} \quad &\sum_{m=1}^M \lVert \mathbf{w}_{m} \rVert_2^2 \\
\text{subject to} \quad & \mathbf{x}_{m, k} - \mathbf{w}_m = 0,~\forall k \in \mathcal{K}, m \in \mathcal{M}, \\
& \mathbf{y}_{m, n} - \mathbf{w}_m = 0,~\forall n \in \mathcal{N}, m \in \mathcal{M}, \\
& \gamma_k \left ( \sum_{j \neq m} \lvert \mathbf{h}_k^H \mathbf{x}_{j, k} \rvert^2 + \sigma_k^2 \right ) - 2 \mathfrak{Re} \left \{ (\mathbf{w}_m^{(t)})^H \mathbf{h}_k \mathbf{h}_k^H \mathbf{x}_{m, k} \right \} \nonumber \\
&\quad + \lvert \mathbf{h}_k^H \mathbf{w}_m^{(t)} \rvert^2 \leq  0, ~\forall k \in \mathcal{G}_m, \forall m \in \mathcal{M}, \label{eqn:CCP-consensus-admm-SINR-constraint}  \\
& \sum_{m=1}^M \mathbf{y}_{m, n}^H \mathbf{R}_n \mathbf{y}_{m, n} \leq P_n,~\forall n \in \mathcal{N}.  \label{eqn:CCP-consensus-admm-perantenna-constraint}
\end{align} \label{eqn:CCP-consensus-admm-subproblem}
\end{subequations}
where $\mathbf{x}_k \triangleq \{ \mathbf{x}_{m, k}  \in \mathbb{C}^N | m \in \mathcal{M}\}$ is a set of auxiliary variables for each SINR constraint $k \in \mathcal{K}$ and $\mathbf{y}_n \triangleq \{ \mathbf{y}_{m, n} \in \mathbb{C}^N | m \in \mathcal{M}\}$ is a set of auxiliary variables for each per-antenna power constraint $n \in \mathcal{N}$.
Then the corresponding ADMM algorithm is summarized in Alg. \ref{alg:QoS-subproblem-consensus-ADMM}.
\begin{algorithm}[h]
\caption{Consensus-ADMM for solving problem $\mathcal{Q}^{(t)}$} \label{alg:QoS-subproblem-consensus-ADMM}
\begin{algorithmic}[0]
\STATE \textbf{Initialization:} Initialize $\mathbf{w}_{m}^{0} \leftarrow \mathbf{w}_{m}^{(t)}, \mathbf{u}_{m, k}^{0} \leftarrow \mathbf{0}, \mathbf{v}_{m, n}^{0} \leftarrow 0, \forall m \in \mathcal{M}, k \in \mathcal{K}, n \in \mathcal{N}$, and $j \leftarrow 0$. Set the penalty parameter $\rho$.
\STATE \textbf{Repeat}
	\begin{enumerate}
	\item Update the first block of variables $\{\mathbf{x}_k^{j+1}, \mathbf{y}_n^{j+1}, \forall k \in \mathcal{K}, n \in \mathcal{N}\}$
	\begin{align}
	\mathbf{x}_{k}^{j+1} := \arg\min_{\mathbf{x}_{k}} &\sum_{m=1}^M \lVert \mathbf{x}_{m, k} - \mathbf{w}_m^{j} + \mathbf{u}_{m, k}^{j}\rVert^2  \label{eqn:consensus-admm-x-update-subproblem} \\
	\text{s.t.}~ &\gamma_k \left ( \sum_{m \neq m_k} \lvert \mathbf{h}_k^H \mathbf{x}_{m, k} \rvert^2 + \sigma_k^2 \right ) - 2 \mathfrak{Re} \left \{ (\mathbf{w}_{m_k}^{(t)})^H \mathbf{h}_k \mathbf{h}_k^H \mathbf{x}_{m_k, k} \right \} \nonumber \\
	&\quad + \lvert \mathbf{h}_k^H \mathbf{w}_{m_k}^{(t)} \rvert^2 \leq  0, \nonumber \\
	\mathbf{y}_{n}^{j+1} := \arg\min_{\mathbf{y}_{n}} & \sum_{m=1}^M \lVert \mathbf{y}_{m, n} - \mathbf{w}_m^{j} + \mathbf{v}_{m, n}^{j}\rVert_2^2  \label{eqn:consensus-admm-y-update-subproblem}\\
	\text{s.t.}~  &\sum_{m=1}^M \mathbf{y}_{m, n}^H \mathbf{R}_n \mathbf{y}_{m, n} \leq P_n. \nonumber
	\end{align}
	\item Update the second block of variables $\mathbf{w}^{j+1}$
	\begin{align} \label{eqn:consensus-admm-W-update-solution}
	\mathbf{w}_{m}^{j+1} := \frac{\rho}{2 + \rho (K + N)}  \left ( \sum_{k}  (\mathbf{x}^{j+1}_{m, k} + \mathbf{u}_{m, k}^j ) + \sum_{n} (\mathbf{y}^{j+1}_{m, n} + \mathbf{v}_{m, n}^{j} )\right), \forall m \in \mathcal{M}.
	\end{align}
	\item Update the dual variables $\{\mathbf{u}_k^{j+1}, \mathbf{v}_n^{j+1}, \forall k \in \mathcal{K}, n \in \mathcal{N}\}$
	\begin{align}
	\mathbf{u}_{m, k}^{j+1} &:= \mathbf{u}_{m, k}^{j} + \mathbf{x}_{m, k}^{j+1} - \mathbf{w}_m^{j+1},~\forall m \in \mathcal{M}, \\
	\mathbf{v}_{m, n}^{j+1} &:= \mathbf{v}_{m, n}^{j} + \mathbf{y}_{m, n}^{j+1} -\mathbf{w}_m^{j+1},~\forall m \in \mathcal{M}.
	\end{align}
   \item Set $j \leftarrow j + 1$.
\end{enumerate}
\STATE \textbf{Until} convergence criterion is met.
\end{algorithmic}
\end{algorithm}

The update of variable $\mathbf{x}_k$ for each $k \in \mathcal{K}$ in problem \eqref{eqn:consensus-admm-x-update-subproblem} and $\mathbf{y}_n$ for each $n \in \mathcal{N}$ in problem \eqref{eqn:consensus-admm-y-update-subproblem} are convex QCQP-1 problems, which are similar to the $\mathbf{\Gamma}$ update in Section \ref{sec:CCP-ADMM_Gamma_update} and $\mathbf{v}$ update in Section \ref{sec:CCP-ADMM_v_update}, respectively. The details are therefore omitted here.

\section{Finding a closed-form solution of Problem \eqref{eqn:FPP-admm-Gamma-update-subproblem}} \label{solution-for-FPP-admm-Gamma-update-subproblem}
The Lagrangian of problem \eqref{eqn:FPP-admm-Gamma-update-subproblem} is given by
\begin{align}
\mathcal{L}_{\pi} =& \sum_{m=1}^M \lvert \Gamma_{k, m} - \mathbf{h}_k^H \mathbf{w}_m^{j} + \lambda_{k,m}^{j}\rvert_2^2 + \pi  \left [ \gamma_k \left ( \sum_{m \neq m_k} \lvert \Gamma_{k, m} \rvert^2 + \sigma_k^2 \right ) - \lvert \Gamma_{k, m_k} \rvert^2  \right ], \label{Lagrangian-FPP-admm-Gamma-update-subproblem}
\end{align}
where $\pi \geq 0$ is the dual variable associated with the inequality constraint \eqref{eqn:FPP-admm-Gamma-update-constraint}. If problem \eqref{eqn:FPP-admm-Gamma-update-subproblem} is feasible, then from the dual of the problem, there should be $\pi \leq 1$ \cite[Appendix B]{Boyd_convex_optimization}. Setting the derivative of $\mathcal{L}_{\pi}$ with respect to $\{\Gamma_{k,m} \}_{m=1}^M$ to zero, we have
\begin{align}  \label{equ:FPP-admm-derivative-solution}
\begin{cases}
	(\pi \gamma_k + 1) \Gamma_{k, m} = \mathbf{h}_k^H \mathbf{w}_m^{j} - \lambda_{k,m}^{j}  ,&\text{ if } m \neq m_k, m \in \mathcal{M}, \\
	(1 - \pi) \Gamma_{k, m} = \mathbf{h}_k^H \mathbf{w}_{m}^{j} - \lambda_{k,m}^{j}, &\text{ if } m = m_k.
\end{cases}
\end{align}
If the optimal dual variable $\pi^*$ satisfies $\pi^* = 1$, we must have $\mathbf{h}_k^H \mathbf{w}_{m_k}^{j} - \lambda_{k,m_k}^{j} = 0$. Then $\{\Gamma_{k,m} \}_{m=1}^M$ can be obtained from \eqref{equ:FPP-admm-derivative-solution}. Otherwise, we have
\begin{align}  \label{equ:FPP-admm-Gamma-solution}
\Gamma_{k, m} =
\begin{cases}
	\left( \mathbf{h}_k^H \mathbf{w}_m^{j} + \lambda_{k,m}^{j} \right ) / (\pi \gamma_k + 1) ,&\text{ if } m \neq m_k, m \in \mathcal{M}, \\
	\left( \mathbf{h}_k^H \mathbf{w}_m^{j} + \lambda_{k,m}^{j} \right ) / (1 - \pi), &\text{ if } m = m_k.
\end{cases}
\end{align}
Substituting the solution \eqref{equ:FPP-admm-Gamma-solution} back into the inequality constraint \eqref{eqn:FPP-admm-Gamma-update-constraint}, we have
\begin{align}
\tilde{g}(\pi) = &  \frac{ \gamma_k \sum_{m \neq m_k} \lvert \mathbf{h}_k^H \mathbf{w}_m^{j} - \lambda_{k,m}^{j} \rvert^2} {(\pi \gamma_k + 1)^2} + \gamma_k\sigma_k^2 -  \frac{ \lvert \mathbf{h}_k^H \mathbf{w}_{m_k}^{j} - \lambda_{k,{m_k}}^{j} \rvert^2} {(1-\pi)^2} \leq 0. \nonumber
\end{align}

It is easy to verify that $\nabla \tilde{g}(\pi) < 0$ and $\tilde{g}(\pi)$ is monotonically decreasing in the region $0 \leq \pi < 1$. According to the complementary slackness condition, we have $\pi^* \tilde{g}(\pi^*) = 0$. Therefore, if $\tilde{g}(0) < 0$, we have $\pi^* = 0$. Otherwise, $\tilde{g}(\pi)$ has a zero-crossing in the region $0 \leq \pi < 1$ and we have $\tilde{g}(\pi^*) = 0$. For ease of notation, we let
\begin{align}
&\tilde{a} \triangleq \gamma_k  \sum_{m \neq m_k} \lvert \mathbf{h}_k^H \mathbf{w}_m^{j} - \lambda_{k,m}^{j} \rvert^2, \nonumber \\
&\tilde{b} \triangleq \gamma_k \sigma_k^2, \nonumber \\
&\tilde{c} \triangleq \lvert \mathbf{h}_k^H \mathbf{w}_{m_k}^{j} - \lambda_{k,{m_k}}^{j} \rvert^2. \nonumber
\end{align}
Note that $\tilde{g}(\pi) = 0$ is equivalent to solving the quartic equation
\begin{align*}
\tilde{h}(\pi) = & (\tilde{b}\gamma_k^2)\pi^4 + (2 \tilde{b} \gamma_k - 2 \tilde{b} \gamma_k^2)\pi^3 + (\tilde{b} \gamma_k^2 + \tilde{b} - 4 \tilde{b} \gamma_k + \tilde{a} - \tilde{c} \gamma_k^2) \pi^2 \\
& \quad + (2 \tilde{b} \gamma_k - 2 \tilde{b} -2 \tilde{a} -2 \tilde{c} \gamma_k)\pi + (\tilde{b} + \tilde{a} - \tilde{c} ) = 0,
\end{align*}
of which the closed-form solution can be expressed using the quartic formula. The root satisfying $0 \leq \pi < 1$ is guaranteed to be the unique (thus correct) solution for $\pi^*$. Then, the optimal $\{\Gamma_{k,m}\}_{m=1}^M$ can be recovered through equation \eqref{equ:FPP-admm-Gamma-solution} with the obtained $\pi^*$.

\section{Proof of Lemma \ref{lemma_feasible_point}} \label{proof_of_lemma_feasible_point}
To find a point that satisfies the SINR constraints, we propose to solve the following problem with the objective of minimizing $\sum_{k=1}^K \sum_{m=1}^M \lvert \mathbf{h}_k^H \mathbf{w}_m\rvert^2$ subject to the SINR constraints
\begin{align} \label{eqn:lemma1-problem}
\mathop{\text{minimize}}_{\mathbf{w}} \quad & \sum_{k=1}^K \sum_{m=1}^M \lvert \mathbf{h}_k^H \mathbf{w}_m\rvert^2 \\
\text{subject to}\quad & \gamma_k \left ( \sum_{j \neq m} \lvert \mathbf{h}_k^H \mathbf{w}_j \rvert^2 + \sigma_k^2 \right ) - \lvert \mathbf{h}_k^H \mathbf{w}_m \rvert^2 \leq  0, ~\forall k \in \mathcal{G}_m, \forall m \in \mathcal{M}. \nonumber
\end{align}
Let $A_{k, m} = \mathbf{h}_k^H \mathbf{w}_m,~\forall k \in \mathcal{K}, m \in \mathcal{M}$, then problem \eqref{eqn:lemma1-problem} can be decomposed into $K$ subproblems, one for each $k \in \mathcal{K}$:
\begin{align} \label{eqn:lemma1-subproblem}
\mathop{\text{minimize}}_{\{A_{k,m}\}_{m=1}^M} \quad & \sum_{m=1}^M \lvert A_{k, m} \rvert^2 \\
\text{subject to}\quad & \gamma_k \left ( \sum_{m \neq m_k} \lvert A_{k, m} \rvert^2 + \sigma_k^2 \right ) - \lvert  A_{k, m_k} \rvert^2 \leq  0. \nonumber
\end{align}
where $m_k$ is the index of the multicast group that user $k$ belongs to. Each subproblem \eqref{eqn:lemma1-subproblem} is a QCQP-1 problem, the optimal solution of which is given by
\begin{equation}  \label{eqn:solution-lemma1-subproblem}
A_{k, m} =
	\begin{cases}
	\sqrt{\gamma_k \sigma_k^2} e^{j\theta_{k}}, &\text{ if } m = m_k,  \\
	0, &\text{ otherwise},
	\end{cases}
\end{equation}
where $\theta_{k} \in [0, 2\pi]$ is arbitrary. Gathering the solution \eqref{eqn:solution-lemma1-subproblem} for all $k \in \mathcal{K}$, we obtain the equation \eqref{eqn:solution-A}.

Since the channel matrix $\mathbf{H}$ has full column rank, the equation $\mathbf{h}_k^H \mathbf{w}_m = A_{k,m}, \forall k,m$, or equivalently, $\mathbf{H}^H \mathbf{W} = \mathbf{A}$ always holds with $\mathbf{W} = \mathbf{H}(\mathbf{H}^H\mathbf{H})^{-1} \mathbf{A}$, where $\mathbf{A}$ is a $K \times M$ matrix with the $(k, m)$-th element defined as $ A_{k,m}$.
Note that the objective function $\sum_{k=1}^K \sum_{m=1}^M \lvert \mathbf{h}_k^H \mathbf{w}_m\rvert^2$ in problem \eqref{eqn:lemma1-problem} is a key to enable the simple and closed-form solution \eqref{eqn:solution-lemma1-subproblem} for each subproblem \eqref{eqn:lemma1-subproblem}. Replacing it with a different objective function may not result in closed-form solution.

\bibliographystyle{IEEEtran}
\bibliography{IEEEabrv,admm_multicast}

\end{document}